\documentclass[nofootinbib,superscriptaddress,citeautoscript,aps,jcp,groupedaddress,10pt]{revtex4-2}

\usepackage{graphicx}
\usepackage{amsmath,amssymb}
\usepackage[dvipsnames]{xcolor}
\usepackage{float}
\usepackage{listings}
\usepackage{romannum}

\newcommand{\beq}{\begin{eqnarray}}
\newcommand{\eeq}{\end{eqnarray}}
\def\beqa{\begin{eqnarray}}
\def\eeqa{\end{eqnarray}}

\begin{document}
\pagenumbering{arabic}
\author{Yael Avni$^1$}
\author{David Andelman$^{1}$\footnote{corresponding author, email: andelman@post.tau.ac.il}}
\author{Henri Orland$^{2}$}
\affiliation{${}^{1}$School of Physics and Astronomy, Tel Aviv University, Ramat Aviv 69978, Tel Aviv, Israel}
\affiliation{${}^{2}$Institut de Physique Th\'eorique, Universit\'e de Paris-Saclay, CEA, CNRS, F-91191 Gif-sur-Yvette Cedex, France}
\title{Conductance of concentrated electrolytes: multivalency and the Wien effect}
\begin{abstract}
The electric conductivity of ionic solutions is well understood at low ionic concentrations of up to a few millimolar but becomes difficult to unravel at higher concentrations that are still common in nature and technological applications. A model for the conductivity at high concentrations was recently put forth for monovalent electrolytes at low electric fields. The model relies on applying a stochastic density-functional theory and using a modified electrostatic pair-potential that suppresses unphysical, short-range electrostatic interactions. Here, we extend the theory to multivalent ions as well as to high electric fields where a deviation from Ohm's law known as the {\it Wien effect} occurs. Our results are in good agreement with experiments and recent simulations.
\end{abstract}
\maketitle

\section{Introduction} \label{Introduction}
Understanding the electric conductance of concentrated electrolytes has posed a great theoretical challenge for over a century. The theory of electrolytic conductivity was pioneered by Debye and H\"uckel~\cite{DH}. They used the notion of an {\it ionic cloud}, where each ion is assumed to be surrounded by a smeared ionic distribution of net opposite charge, which gets distorted upon movement of the central ion. Onsager detected a flaw in the Debye-H\"uckel account of the central ion diffusion~\cite{Collected} and a few years later corrected the theory, yielding the so-called Debye-H\"uckel-Onsager (DHO) equation (also known as the ``Onsager limiting law") for the conductivity of electrolytes~\cite{Onsager,Onsager2}. Due to its elegance and accurate predictions, the DHO equation is considered to be one of the cornerstones of electrolyte theory.

A few decades after it was established, the DHO equation was extended to arbitrary electric-field strengths by Onsager and Kim~\cite{Onsager1957}, who relied on the unpublished thesis of Wilson on binary electrolytes~\cite{Wilson1936}. The modified theory (often called ``Onsager-Wilson (OW) theory'') captures the {\it Wien effect}~\cite{Wien1, Wien2, Eckstrom1939}, which is an increase in the conductivity with the electric-field strength, attributed to the destruction of the ionic cloud. A related phenomenon called ``the second Wien effect"~\cite{Onsager1934,Kaiser2013}, occurs in weak electrolytes, where the conductivity increases with the electric field strength due to a modification in the dissociation kinetics of bound pairs.

While being a remarkable achievement, the DHO and OW theories can be applied only for very dilute electrolyte solutions. They break down when the ion concentration exceeds the threshold of a few millimolar for monovalent ions and an even lower threshold for multivalent ions~\cite{Bockris,RobinsonStokes}. Since its onset in the 1920s', there have been many attempts to extend the DHO theory to higher concentrations. Initially, by Onsager himself (the ``Onsager-Fuoss" theory)~\cite{OnsagerFuoss1957,OnsagerFuoss1962}, and later on by others~\cite{Pitts1953,Friedman1983,Bernard1991,Chandra1999,Fraenkel2018,Zhang2020,Nagele1,Nagele2}. However, previous studies either used fit parameters that limit their predictive power or included very complicated results that are difficult to use and are not thoroughly transparent. Moreover, to the best of our knowledge, no previous work in the concentrated regime was generalized to finite electric-field strengths; Thus, not capturing the Wien effect.

In recent years, highly concentrated electrolytes have attracted a lot of attention~\cite{Kornyshev2020,Feng2019,Adar2019,Benaglia2021} due to their numerous potential applications and surprising experimental observations~\cite{Israelachvili2013,Perkin2016,Perkin2017}. At the same time, advances in nonequilibrium theories such as stochastic density functional theory (often referred to as the Kawasaki-Dean equation)~\cite{Kawasaki1994,Dean1996,Vrugt2020,Golestanian2018}, have led to a new way of calculating the ionic conductivity in the dilute limit~\cite{Demery2016,Peraud2017,Donev2019}, which is far simpler than the previous ionic-cloud-based approach. Relying on these theoretical advances and using a modified pair-potential to account for the finite ion size, we recently formulated~\cite{Avni2022} a new model for the conductivity of concentrated electrolytes. The model was shown to agree well with experimental data for different aqueous solutions at concentrations as high as $3$\,M but was limited to binary monovalent ions and low electric fields.

In the present work, we extend this model and apply it to multivalent ions and finite electric fields. We derive a general expression for the conductivity of binary electrolytes and then focus on two cases: (i) the weak-field limit with any $z_1{:}z_2$ ionic valencies, and; (ii) the symmetric $z{:}z$ electrolyte at finite field intensities, where we recover the Wien effect and provide new predictions for the high concentration regime. Our results compare favorably to experiments and recent simulations.

The outline of this paper is as follows: in Sec.~\ref{model}, we present the model system and derive the conductivity equations for ionic solutions with an arbitrary number of species, at high ionic concentrations. In Sec.~\ref{Two}, we restrict ourselves to binary electrolytes and analyze the low electric-field limit as well as the case of symmetric ions with any finite electric field. In Sec.~\ref{Comparison}, we compare our results to experiments and simulations. Finally, in Sec.~\ref{Conclusions}, we conclude and suggest future experiments to further test our predictions.

\section{The model} \label{model}
\subsection{The Equations of Motion} \label{formalism}
We consider a homogeneous ionic solution composed of $M$ ionic species of charge ${q_\alpha}$ and average concentration $n^0_{\alpha}$, where ${\alpha=1,...,M}$. The ions are embedded in a solvent with dielectric permittivity $\varepsilon$ and viscosity $\eta$ at temperature $T$. The solution is subjected to a constant (static) external electric field ${\boldsymbol{E}_{0}}$ pointing in a fixed direction.

The local ionic concentrations, denoted by $n_\alpha({\bf r},t)$, satisfy the continuity equation
\beq \label{continuity}
\partial_{t}n_{\alpha} = -\boldsymbol{\nabla}\cdot\boldsymbol{j}_{\alpha}\,\,\,\,\,\,\,\,\,\,\,\, \alpha={1,...,M},
\eeq
where ${\boldsymbol j_\alpha}({\boldsymbol r},t)$ is the ionic flux of the $\alpha$ species, given by,
\beq \label{j}
\boldsymbol{j}_{\alpha}=n_{\alpha}\boldsymbol{u}-D_{\alpha}\boldsymbol{\nabla}n_{\alpha}+\mu_{\alpha}\boldsymbol{f}_{\alpha}-\sqrt{2D_{\alpha}n_{\alpha}}\boldsymbol{\zeta}_{\alpha}.
\eeq
The first and second terms on the right-hand-side of Eq.~(\ref{j}) are advection and diffusion terms, respectively, where $\boldsymbol{u}(\boldsymbol{r},t)$ is the solvent velocity field, and $D_\alpha$ is the diffusion coefficient of the $\alpha$ species at infinite ionic dilution. The third term accounts for the motion due to the external field and inter-ionic forces. Here, $\mu_\alpha$ is the ion mobility at infinite ionic dilution, related to $D_{\alpha}$ by the Einstein relation $\mu_\alpha=D_\alpha/k_BT$, with $k_B$ being the Boltzmann constant, and $\boldsymbol{f}_{\alpha}(\boldsymbol{r},t)$ is the force density given by
\beq \label{force_density}
\boldsymbol{f}_{\alpha}=n_{\alpha}q_{\alpha}\boldsymbol{E}_{0}-n_{\alpha}\sum_{\beta=1}^M\int{\rm d}^{3}r'\,n_{\beta}\left(\boldsymbol{r}',t\right)\boldsymbol{\nabla}v_{\alpha \beta}\left(\left|\boldsymbol{r}-\boldsymbol{r}'\right|\right),
\eeq
where $v_{\alpha \beta}$ is the pair interaction energy between ions of species $\alpha$ and $\beta$. Note that for generality sake we do not specify $v_{\alpha \beta}$ until Sec.~\ref{modified}. The last term in Eq.~(\ref{j}) is a stochastic flux, where $\boldsymbol{\zeta}_{\alpha}(\boldsymbol{r},t)$ is a 3D white-noise function, satisfying
\beqa
&& \langle\boldsymbol{\zeta}_{\alpha}\left(\boldsymbol{r},t\right)\rangle =0\\
&&\langle{\zeta}_{\alpha}^{n}\left(\boldsymbol{r},t\right){\zeta}_{\beta}^{m}\left(\boldsymbol{r}',t'\right)\rangle = \delta_{\alpha \beta}\delta_{nm}\delta\left(t-t'\right)\delta\left(\boldsymbol{r}-\boldsymbol{r}'\right),\nonumber
\eeqa
where $n$ and $m$ denote the cartesian components of the vector $\boldsymbol{\zeta}_{\alpha}$. Equations~(\ref{continuity}) and~(\ref{j}) can be derived by transforming the Langevin equation from individual particle representation to concentration fields using Ito calculus, and it is referred to as {\it stochastic density-functional theory} (SDFT)~\cite{Dean1996}.\footnote{The derivation of the stochastic density-functional theory in Ref.~\cite{Dean1996} was done without considering the advection by the solvent. However, advection can be easily incorporated into the formalism by adding the solvent velocity to the Langevin equation, yielding Eqs.~(\ref{continuity}) and~(\ref{j}) exactly.}

The ion continuity equation is coupled to the Navier-Stokes equation for an incompressible fluid,
\beqa \label{Stokes0}
&&\rho \left[ \partial_t\boldsymbol{u}+(\boldsymbol{u}\cdot\nabla)\boldsymbol{u}\right]=\eta\nabla^{2}\boldsymbol{u}-\boldsymbol{\nabla}p+\sum_{\alpha=1}^M \boldsymbol{f}_{\alpha}
\eeqa
where $p(\boldsymbol{r},t)$ is the local pressure and $\rho$ is the solvent density. The $\rho(\boldsymbol{u}\cdot\nabla)\boldsymbol{u}$ term will disappear when we linearize the equations of motion around $\boldsymbol{u}=0$ in the next~\ref{conductivity_calc} subsection. The $\rho\partial_t\boldsymbol{u}$ term can also be neglected due to the following argument. As long as the electric field is not too strong, the typical time and length scales that characterize the ionic motion, $T$ and $L$, satisfy $L^2/T\simeq D_{\alpha}$. Applying this rescaling we get that
$|\rho\partial_t \delta \boldsymbol{u}|/|\eta\nabla^2\boldsymbol{u}| \simeq \rho D_{\alpha}/\eta$, which is the inverse of the Schmidt number~\cite{Bergman2011} and is roughly $\sim 0.001$ for standard electrolytes. Therefore, the resulting equation for the solvent velocity is the Stokes equation for an incompressible fluid,
\beqa \label{Stokes2}
&&\boldsymbol{\nabla}\cdot \boldsymbol{u}=0\nonumber\\
&&\eta\nabla^{2}\boldsymbol{u}-\boldsymbol{\nabla}p+\sum_{\alpha=1}^M \boldsymbol{f}_{\alpha}=0.
\eeqa
%

\subsection{Calculation of the conductivity} \label{conductivity_calc}

The conductivity of the ionic solution is defined by the ratio
\beq \label{kappa_def}
\kappa=\langle J_{\parallel}\rangle/E_{0},
\eeq
where  $\langle ... \rangle$ is the thermodynamic ensemble average,
and $\boldsymbol{J}(\boldsymbol{r},t)$ is the electric current density, which depends on the ionic fluxes, $\boldsymbol{j}_{\alpha}$,
\beq \label{J}
\boldsymbol{J}=\sum_{\alpha=1}^M q_\alpha\boldsymbol{j}_{\alpha}(\boldsymbol{r},t).
\eeq
The subscript ``$\parallel$" in Eq.~(\ref{kappa_def}) denotes the vector projection on the external field direction, $J_{\parallel}={\boldsymbol J}\cdot\hat{ E_0}$. Substituting Eq.~(\ref{j}) in Eq.~(\ref{J}) and performing the average in Eq.~(\ref{kappa_def}), we obtain
\beqa \label{Full_kappa1}
&& \kappa = \kappa_{0}+\kappa_{\text{hyd}}+\kappa_{\text{el}} \nonumber\\
&&  \kappa_{\text{hyd}} =\sum_\alpha \frac{q_\alpha}{E_{0}} \langle u_{\parallel}\left(\boldsymbol{r},t\right) n_\alpha \left(\boldsymbol{r},t\right)\rangle  \nonumber\\
&&  \kappa_{\text{el}} = -\sum_{\alpha,\beta}\frac{q_{\alpha} \mu_{\alpha}}{E_{0}}\int{\rm d}^{3}{r}'\,  \partial_{r_\parallel}v_{\alpha \beta}\left(\left|\boldsymbol{r}-\boldsymbol{r}'\right|\right) \langle n_{\alpha}\left(\boldsymbol{r},t\right)n_\beta\left(\boldsymbol{r}',t\right)\rangle,
\eeqa
where $\kappa_{0}$ is the conductivity at infinite dilution,
\beq \label{kappa_0}
\kappa_{0}=\sum_{\alpha=1}^M q^2_\alpha \mu_\alpha n^0_{\alpha}.
\eeq
Note that $\kappa_0$ depends linearly on the concentrations as the ions do not interact in this limit. In order to obtain Eq.~(\ref{Full_kappa1}), we invoke the system homogeneity and the independence between $n_\alpha$ and ${\boldsymbol \zeta}_\alpha$ at equal times.

Equation~(\ref{Full_kappa1}) implies that at finite concentrations the conductivity deviates from its dilute-limit behavior due to two effects. The first effect, incorporated in $\kappa_{\text{hyd}}$, is a hydrodynamically mediated interaction between the ions, and it is traditionally referred to as the {\it electrophoretic effect}. We note that in its present form, the average $\langle u_{\parallel}\left(\boldsymbol{r},t\right) n_\alpha \left(\boldsymbol{r},t\right)\rangle$ in $\kappa_{\text{hyd}}$ includes the ion interaction with its own induced velocity field, resulting in a self-interaction that should be subtracted, as will be done later on. The second effect, incorporated in $\kappa_{\text{el}}$, is a direct ionic interaction that is mostly electrostatic but will include finite-size corrections as we explain in Sec.~\ref{modified}, below. This effect is often referred to as the {\it relaxation effect}.

In order to calculate the averages in Eq.~(\ref{Full_kappa1}) we need to solve the equations of motion, Eqs.~(\ref{continuity}), (\ref{j}) and~(\ref{Stokes2}). An exact solution cannot be obtained. Instead, we linearize the equations around the mean field value corresponding to no noise. This is done by writing $n_{\alpha}({\boldsymbol r},t)  =n^0_{\alpha} +\delta n_{\alpha}({\boldsymbol r},t)$, $\boldsymbol{u}({\boldsymbol r},t)  =\delta\boldsymbol{u}({\boldsymbol r},t)$ and $p({\boldsymbol r},t)  =p_0 + \delta{p}({\boldsymbol r},t)$, and keeping only terms up to linear order in $\delta n_{\alpha}$, $\delta \boldsymbol{u}$, $\delta p$, and $\zeta_{\alpha}$. The linearization is justified for small fluctuations around the mean-field values. In Fourier space, the linearized form of the ion equation of motion is
\beq \label{matrix_equation}
\frac{\partial \delta{\tilde{n}}_{\alpha}(\boldsymbol{k},t)}{\partial t}=A_{\alpha \beta}(\boldsymbol{k})\delta\tilde{n}_{\beta}(\boldsymbol{k},t)+B_{\alpha \beta}(\boldsymbol{k})\tilde{\zeta}_{\beta}(\boldsymbol{k},t),
\eeq
where $\tilde{f}(\boldsymbol{k})=\int {\rm d}^3r\, f(\boldsymbol{r}){\rm e}^{-i\boldsymbol{k}\cdot\boldsymbol{r}}$ is the Fourier transform of the function $f(\boldsymbol{r})$. The matrices $A(\boldsymbol{k})$ and $B(\boldsymbol{k})$ are
\beqa \label{matrices}
A_{\alpha\beta}(\boldsymbol{k}) && =\begin{cases}
-D_{\alpha}k^{2}-i\mu_{\alpha}q_{\alpha}k_{\parallel}E_{0}-\mu_{\alpha} n^0_{\alpha} k^{2}\tilde{v}_{\alpha\alpha}(k) & \alpha=\beta\nonumber\\[3pt]
-\mu_{\alpha} n^0_{\alpha} k^{2}\tilde{v}_{\alpha\beta}(k) & \alpha\neq\beta
\end{cases}\\
B_{\alpha\beta}(\boldsymbol{k}) && =\sqrt{2D_{\alpha} n^0_{\alpha}}k\delta_{\alpha\beta}.
\eeqa
Note that we used the fact that $i\boldsymbol{k}\cdot\tilde{\boldsymbol{\zeta}}_{\alpha}(\boldsymbol{k},t)=i\sum\limits_{i=1}^{3} k_i \tilde{\zeta}^{i}_{\alpha}(\boldsymbol{k},t)$, is a sum of three independent white noise functions with zero mean. Therefore, it is a scalar white noise function with zero mean and correlation given by
\begin{eqnarray}
\label{noise}
\langle (i\boldsymbol{k}\cdot\tilde{\boldsymbol{\zeta}}_{\alpha}(\boldsymbol{k},t))\cdot(i\boldsymbol{k'}\cdot\tilde{\boldsymbol{\zeta}}_{\beta}(\boldsymbol{k'},t')) \rangle &=&- \sum\limits_{i=1}^{3}\sum\limits_{j=1}^{3}  k_i  k'_j \langle \tilde{\zeta}^{i}_{\alpha}(\boldsymbol{k},t) \tilde{\zeta}^{j}_{\beta}(\boldsymbol{k'},t')\rangle  \nonumber \\
&=&-\left(2\pi\right)^{3}\sum\limits_{i=1}^{3}\sum\limits_{j=1}^{3} k_i k'_j  \delta_{ij} \delta_{\alpha \beta}\delta(t-t')\delta(\boldsymbol{k}+\boldsymbol{k'}) \nonumber \\
&=&\left(2\pi\right)^{3}k^2 \delta_{\alpha \beta} \delta(t-t')\delta(\boldsymbol{k}+\boldsymbol{k'}) 
\end{eqnarray}
which can be represented as $k \tilde {\zeta}_{\alpha}(\boldsymbol{k},t)$, where $\tilde{\zeta}_{\alpha}(\boldsymbol{k},t)$ is a white-noise scalar function with zero mean and correlation given by ${\langle\tilde{\zeta}_{\alpha}(\boldsymbol{k},t)\tilde{\zeta}_{\beta}(\boldsymbol{k}',t')\rangle =\left(2\pi\right)^{3}\delta_{\alpha \beta}\delta\left(t-t'\right)\delta\left(\boldsymbol{k}+\boldsymbol{k}'\right)}$.

The linearized form of the Stokes equation in Fourier space is
\beq \label{Stokes_Fourier}
k^{2}\eta\delta\tilde{\boldsymbol{u}}\left(\boldsymbol{k}\right)=-i\boldsymbol{k}\delta\tilde{p}\left(\boldsymbol{k}\right)+\sum_{\alpha}q_{\alpha}\boldsymbol{E}_{0}\delta\tilde{n}_{\alpha}\left(\boldsymbol{k}\right)+i\boldsymbol{k}\sum_{\alpha,\beta}n^0_{\alpha}\,\tilde{v}_{\alpha\beta}(k)\delta\tilde{n}_{\beta}(\boldsymbol{k}).
\eeq
We use the Fourier transform of the incompressiblity condition (Eq.~(\ref{Stokes2})), $\boldsymbol{k}\cdot\delta\boldsymbol{\tilde{u}}(\boldsymbol{k})= 0$, to eliminate $\tilde{p}(\boldsymbol{k})$ in Eq.~(\ref{Stokes_Fourier}), and obtain $\delta\tilde{u}_{\parallel}(\boldsymbol{k})$ in terms of the concentrations,
\beq \label{velocity_conc}
\delta\tilde{u}_{\parallel}\left({\boldsymbol k}\right)=\frac{E_{0}}{\eta}\frac{1}{k^{2}}\left(1-\frac{k_{\parallel}^{2}}{k^{2}}\right)\sum_{\alpha=1}^M q_{\alpha}\delta\tilde{n}_{\alpha}\left({\boldsymbol k}\right).
\eeq
Writing Eq.~(\ref{Full_kappa1}) in terms of the fluctuational variables in Fourier space and using Eq.~(\ref{velocity_conc}), we obtain
\beqa
&& \kappa_{\text{hyd}}=\sum_{\alpha,\beta}\frac{q_{\alpha}q_{\beta}}{\eta}\int\frac{{\rm d}^{3}k\,{\rm d}^{3}k'}{\left(2\pi\right)^{6}}{\rm e}^{i\left(\boldsymbol{k}+\boldsymbol{k}'\right)\cdot\boldsymbol{r}}\frac{1}{k^{2}}\left(1-\frac{k_{\parallel}^{2}}{k^{2}}\right)\langle\delta\tilde{n}_{\beta}\left({\boldsymbol{k}},t\right)\delta\tilde{n}_{\alpha}\left(\boldsymbol{k}',t\right)\rangle\nonumber
\\
 && \label{cor2} \kappa_{\text{el}}  = -\sum_{\alpha,\beta}\frac{q_\alpha\mu_{\alpha}}{E_{0}}\int \frac{{\rm d}^{3}k\,{\rm d}^{3}k'}{\left(2\pi\right)^{6}}{\rm e}^{i\left(\boldsymbol{k}+\boldsymbol{k}'\right)\cdot\boldsymbol{r}}\left(ik_{\parallel}'\right)\tilde{v}_{\alpha\beta}({k'})\langle\delta\tilde{n}_{\alpha}(\boldsymbol{k},t)\delta\tilde{n}_{\beta}(\boldsymbol{k}',t)\rangle.
\eeqa
In steady state, the set of linear equations in Eq.~(\ref{matrix_equation}) leads to
\beq \label{correlator_k}
\langle\delta\tilde{n}_{\alpha}(\boldsymbol{k},t)\delta\tilde{n}_{\beta}(\boldsymbol{k}',t)\rangle=\left(2\pi\right)^{3}C_{\alpha \beta}(\boldsymbol{k})\delta\left(\boldsymbol{k}+\boldsymbol{k}'\right),
\eeq
where the correlation matrix, $C(\boldsymbol{k})$, is given by the relation~\cite{RobertZwanzig}
\beq \label{mat_eq}
A(\boldsymbol{k})C(\boldsymbol{k})+C(\boldsymbol{k})A^{\dagger}(\boldsymbol{k})=-B(\boldsymbol{k})B^{\dagger}(\boldsymbol{k}),
\eeq
where $\dagger$ is the Hermitian conjugate. In order to subtract the ion self-correlation, we define the following {\it subtracted correlation matrix} (see supplementary material in Ref.~\cite{Avni2022}):
\beq \label{norm}
\widehat{C}_{\alpha \beta}(\boldsymbol{k})= C_{\alpha \beta}(\boldsymbol{k})-n^0_{\alpha} \delta_{\alpha \beta},
\eeq
and use $\widehat{C}$ instead of $C$ from here on.

Substituting the subtracted correlation matrix in Eq~(\ref{cor2}) and recalling that $\kappa_{{\rm hyd}}$, $\kappa_{{\rm el}}$ and $\tilde{v}_{\alpha\beta}(k)$ are real functions, we obtain
\beqa \label{integrals}
\kappa_{{\rm hyd}}&&=\int\frac{{\rm d}^{3}k}{\left(2\pi\right)^{3}}\frac{1}{\eta k^{2}}\left(1-\frac{k_{\parallel}^{2}}{k^{2}}\right)\sum_{\alpha,\beta}q_\alpha q_\beta\,\text{Re}\left[\widehat{C}_{\alpha\beta}(\boldsymbol{k})\right] \nonumber\\
\kappa_{{\rm el}}&&=-\int\frac{{\rm d}^{3}k}{\left(2\pi\right)^{3}}\:\frac{k_{\parallel}}{E_{0}} \sum_{\alpha,\beta}q_\alpha\mu_\alpha\tilde{v}_{\alpha\beta}({k})\,\text{Im}\left[\widehat{C}_{\alpha\beta}(\boldsymbol{k})\right].
\eeqa
Equation~(\ref{integrals}) shows that the hydrodynamic correction term depends on the real part of the correlation matrix, while the electrostatic term depends on the imaginary part. This can be explained as follows. Let us think of two ions, denoted 1 and 2, and, for convenience, set particle 1 at the origin and particle 2 at position ${\boldsymbol r}$. In Stokes flow, the velocity field generated by the motion of a point-particle is symmetric with respect to its position. Therefore, the velocity that particle 2 induces at the origin, which influences the velocity of particle 1, remains unchanged under the transformation ${\boldsymbol r}
\to-{\boldsymbol r}$. Thus, the hydrodynamic correction depends only on the symmetric part of the correlation function. The force exerted on particle 1 by particle 2 due to the pair-potential $v_{\alpha \beta}(r)$, however, changes its direction under the transformation ${\boldsymbol r}\to -{\boldsymbol r}$. Thus, the electrostatic correction depends on the anti-symmetric part of the correlation function. In Fourier-space, symmetric and anti-symmetric parts of any function transform into real and imaginary parts, respectively, which explains Eq. (\ref{integrals}).

In summary, the correlation matrix is obtained by solving Eq.~(\ref{mat_eq}) and redefining a subtracted correlation matrix $\widehat{C}$ in Eq.~(\ref{norm}). By substituting the matrix $\widehat{C}$ in Eq.~(\ref{integrals}) we can compute the conductivity. 

\subsection{The modified pair-potential} \label{modified}
\begin{figure}
\includegraphics[width = 0.45 \columnwidth,draft=false]{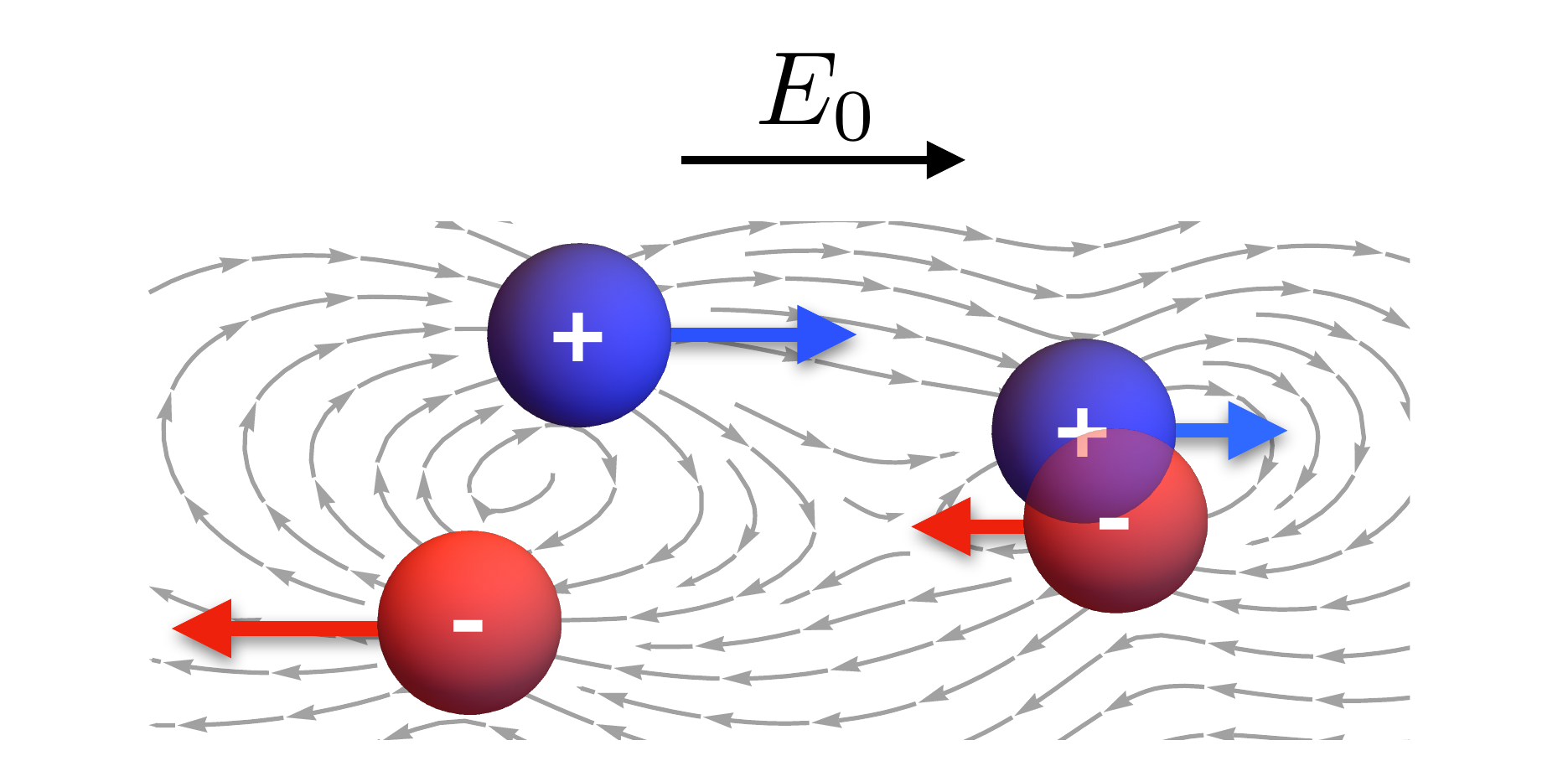} 
\caption{\textsf{A schematic drawing, adapted from Ref.~\cite{Avni2022}, of cations (blue) and anions (red) moving in opposite directions in response to an applied electric field $\boldsymbol{E_0}$. The gray lines represent the fluid velocity field around point-like particles. If the interaction is purely Coulombic, oppositely charged ions are likely to get unrealistically close to one another (right side), thus reducing the conductivity. We use a modified potential to avoid such proximity, prohibited by the ionic finite-size.}}
\label{Fig1}
\end{figure}

Up until now, we did not specify the pair potential, and the results were written in terms of a general $v_{\alpha \beta}(r)$ interaction. For point-like ions, the pair potential equals to the Coulomb interaction, $v_{\alpha \beta}\left(r\right)=q_{\alpha} q_{\beta}/(4\pi\varepsilon_{0}\varepsilon r)$, where $\varepsilon_{0}$ is the vacuum permittivity. However, the point-like approximation leads to an unphysical attraction between oppositely charged ions at distances smaller than the ion diameter (see Fig.~\ref{Fig1}). This becomes problematic at high concentrations, where the ions are more likely to get close to each, leading to an unphysical decrease in the conductivity due to enhanced inter-ionic correlations. This deficiency is present in the DHO theory that assumes point-like ions.

The problem can be remedied by including in $v_{\alpha \beta}(r)$ a hard-core potential,
\beq \label{hardcore}
v_{\alpha\beta}(r)=\begin{cases}
\frac{q_{\alpha}q_{\beta}}{4\pi\varepsilon_{0}\varepsilon r} & r>r_{\alpha}+r_{\beta}\\
\infty & {\rm else}
\end{cases}
\eeq
where $r_\alpha$ is the ion radius, and $r_{\alpha}+r_{\beta}$ is the distance of closest approach between two ions. Unfortunately, such a diverging potential breaks down the perturbative approach introduced in Sec.~\ref{conductivity_calc}. Instead, a viable modification is to introduce a low cutoff to the Coulomb interaction~\cite{Adar2019},
\beq \label{u_co}
v_{\alpha \beta}\left(r\right)=\frac{{q_{\alpha}q_{\beta}}}{4\pi\varepsilon_{0}\varepsilon r}\Theta\left(r-r_\alpha-r_\beta\right),
\eeq
where $\Theta(r)$ is the Heaviside function. In Appendix~\ref{Testing}, Eq.~(\ref{u_co}) is shown to approximate well the average distance between two ions interacting via the pair potential as in Eq.~(\ref{hardcore}), in a confined system that corresponds to concentrated electrolytes (with short inter-ionic distances). While the modified potential presents an uncontrolled approximation, whose range of validity is difficult to define, the comparison made in Appendix~\ref{Testing} supports its validity for aqueous ionic solutions at concentrations of up to roughly $\simeq0.8$~M for monovalent electrolytes, $\simeq0.1$~M for $2{:}1$ electrolytes and $\simeq0.02$~M for $3{:}1$ electrolytes. Furthermore, it shows that the modified potential is far more accurate than the pure Coulomb interaction.

We note that while the modified potential successfully suppresses the short-range electrostatic attraction between oppositely charged ions, it induces an unphysical attraction at very short distances between ions with the same electric charge sign. It might seem that this problem can be circumvented by keeping the standard Coulomb potential (which diverges at small distances) for ions with the same electric charge sign, or by assigning a finite yet large positive value to the potential at $r<r_\alpha+r_\beta$ (a ``soft" repulsion). However, using such potentials is not consistent with the perturbative approach that assumes small interaction energy compared with thermal energy. Thus, in our approach, we keep $v_{\alpha \beta}=0$ for $r<r_\alpha+r_\beta$, for any type of ions, $\alpha$ and $\beta$.

Substituting the Fourier transform of Eq.~(\ref{u_co}), ${\tilde{v}_{\alpha \beta}({k})=q_{\alpha} q_{\beta}\cos\left(kr_{\alpha} + kr_{\beta}\right)/(\varepsilon_{0}\varepsilon k^{2})}$, in Eq.~(\ref{matrices}) and following the analysis of Sec.~\ref{conductivity_calc} leads to a closed-form expression for the conductivity. It is presented in Sec.~\ref{Two} for binary electrolytes.

\section{Binary electrolytes}\label{Two}
We consider hereafter a binary electrolyte $z_+{:}z_-$ containing cations of charge $q_+=ez_+$ and anions of charge $q_-=-ez_-$, where $z_{\pm}$ are the valencies (in absolute value) and $e$ is the elementary charge. The electro-neutrality condition implies $z_+ n^0_+=z_- n^0_-$, where $n^0_{+}$ and $n^0_{-}$ are the average cation and anion concentrations. The experimentally controlled salt concentration, $n_{\rm salt}$, is $n_{\rm salt}\equiv n^0_{+}/z_-=n^0_{-}/z_+$. Note that in the case where the two valencies $z_{\pm}$ have a common divisor ({\it e.g.,} 2:4), it is more natural to define the salt concentration as $n_{\rm salt}$ multiplied by the greatest common divisor. This is done later on for symmetric $z{:}z$ salts.

We make a further simplification by replacing the species-dependent cutoff length in Eq.~(\ref{u_co}), $r_{\alpha}+r_{\beta}$, by a single cutoff that equals the sum of the cation and anion radii, $a\equiv r_+ + r_-$. This simplification is motivated by the fact that the primary role of the cutoff is to prevent attraction between oppositely charged ions. In Appendix~\ref{different cutoffs}, we explore the difference between the conductivity when a single cutoff is used as opposed to three different cutoffs ($r_+ + r_-$, $2r_+$ and $2r_-$), and show that the difference is negligible for standard electrolytes.

Under these simplifications, the conductivity is written as follows:
\beq \label{binary_kappa}
&& \kappa =\kappa_{0}+\kappa_{\text{hyd}}+\kappa_{\text{el}}\nonumber \\
&& \kappa_{{\rm hyd}}=\frac{e^{2}}{\eta}\int\frac{{\rm d}^{3}k}{\left(2\pi\right)^{3}}\frac{1}{k^{2}}\left(1-\frac{k_{\parallel}^{2}}{k^{2}}\right)\left(z_{+}^{2}\widehat{C}_{++}(\boldsymbol{k})+z_{-}^{2}\widehat{C}_{--}(\boldsymbol{k})-2z_{+}z_{-}\,\text{Re}\left[\widehat{C}_{+-}(\boldsymbol{k})\right]\right)\nonumber
\\
&& \label{cond3} \kappa_{{\rm el}} =\frac{e^{3}z_+ z_- (z_+ \mu_++z_-\mu_-)}{E_{0}\varepsilon_{0}\varepsilon}\int\frac{{\rm d}^{3}k}{\left(2\pi\right)^{3}}\:\frac{k_{\parallel}}{k^{2}}\cos\left(ka\right)\text{Im}\left[\widehat{C}_{+-}(\boldsymbol{k})\right],
\eeq
where we used the fact that $\widehat{C}_{\alpha\beta}$ is hermitian. Equation~(\ref{cond3}) indicates that $\kappa_{\rm hyd}$ depends on the difference between the spatial correlations of equal and opposite charges, while $\kappa_{\rm el}$ depends on the spatial correlations between opposite charges only.

The components of the subtracted correlation matrix are 
\beqa \label{C12}
 && \widehat{C}_{\pm\pm}(k)=-\frac{n_{\rm salt} z_+ z_-}{\bar{z}}\frac{\cos\left(ka\right)\left[h^{2}\left(k\right)+2\bar{z}\gamma^{2}\lambda_{D}^{2}l_{E}^{-2}\cos^{2}\theta\left(z_{\mp}\cos\left(ka\right)+2\bar{z}k^{2}\lambda_{D}^{2}\right)\right]}{g\left(k\right)+f\left(k\right)\lambda_{D}^{2}l_{E}^{-2}\cos^{2}\theta}\nonumber\\
\nonumber \\  && \widehat{C}_{+-}(k)=\widehat{C}_{-+}^{\,\ast}(k)=\frac{n_{\rm salt} z_+ z_-}{\bar{z}}\frac{\cos\left(ka\right)\left[h^{2}\left(k\right)-2i\bar{z}\gamma h\left(k\right)\lambda_{D}^{2}l_{E}^{-1}k\cos\theta\right]}{g\left(k\right)+f\left(k\right)\lambda_{D}^{2}l_{E}^{-2}\cos^{2}\theta},
\eeqa
where $\cos \theta={\hat k}\cdot {\hat E}_0$, the Debye screening length is
\beq
\lambda_{D}=\frac{1}{\sqrt{e^{2}(z_{+}^{2}n_{+}^{0}+z_{-}^{2}n_{-}^{0})/\left(\varepsilon\varepsilon_{0}k_{B}T\right)}}=\frac{1}{\sqrt{e^{2}(z_{+}+z_{-})z_{-}z_{+}n_{\rm salt}/\left(\varepsilon\varepsilon_{0}k_{B}T\right)}},
\eeq
and the electric field length is $l_E= k_B T/(e E_0)$ (note that $l_E$ is inversely proportional to the electric-field intensity). We also defined the average valency, $\bar{z}\equiv(z_++z_-)/2$, the parameter $\gamma$,
\beq \label{gamma}
\gamma \equiv \frac{2\left(\mu_{+}z_{+}+\mu_{-}z_{-}\right)}{\left(\mu_{-}+\mu_{+}\right)\left(z_{+}+z_{-}\right)},
\eeq
and the following functions for brevity
\beqa
&&f\left(k\right)=2\gamma^{2}\left[z_{+}\cos\left(ka\right)+2\bar{z}k^{2}\lambda_{D}^{2}\right]\left[z_{-}\cos\left(ka\right)+2\bar{z}k^{2}\lambda_{D}^{2}\right]\nonumber\\
 && g\left(k\right)=2\left[\cos\left(ka\right)+k^{2}\lambda_{D}^{2}\right]\left[\gamma\cos\left(ka\right)+2k^{2}\lambda_{D}^{2}\right]^{2}\nonumber\\
 && h\left(k\right)=\gamma\cos\left(ka\right)+2k^{2}\lambda_{D}^{2}.
\eeqa
For $z_+\neq z_-$, $\gamma<1$ ($\gamma>1$) if the ion with the larger valency has a smaller (larger) mobility. Typically, multivalent ions have smaller mobilities than monovalent ions; thus, asymmetric salts commonly have $\gamma<1$ (see Table~\Romannum{2} in Sec.~\ref{Comparison}). For symmetric salts with $z_+=z_-\equiv z$, $\gamma=1$ and $n^0_+=n^0_-\equiv n$, and the correlation matrix reduces to
\beqa \label{C_symmetric}
 && \widehat{C}_{\pm\pm}(k)=-\frac{n\cos\left(ka\right)\left[\cos\left(ka\right)+2\lambda_{D}^{2}\left(k^{2}+z^{2}l_{E}^{-2}\cos^{2}\theta\right)\right]}{2\left[\cos\left(ka\right)+2k^{2}\lambda_{D}^{2}\right]\left[\cos\left(ka\right)+\lambda_{D}^{2}\left(k^{2}+z^{2}l_{E}^{-2}\cos^{2}\theta\right)\right]}\nonumber\\
 && \widehat{C}_{+-}(k)=\widehat{C}_{-+}^{\,\ast}(k)=\frac{n\cos\left(ka\right)\left[\cos\left(ka\right)+2\lambda_{D}^{2}\left(k^{2}-izl_{E}^{-1}k\cos\theta\right)\right]}{2\left[\cos\left(ka\right)+2k^{2}\lambda_{D}^{2}\right]\left[\cos\left(ka\right)+\lambda_{D}^{2}\left(k^{2}+z^{2}l_{E}^{-2}\cos^{2}\theta\right)\right]}.
\eeqa
We note that $n=z n_{\rm salt}$ as explained at the beginning of Sec.~\ref{Two}.

\begin{figure}
\includegraphics[width = 0.95 \columnwidth,draft=false]{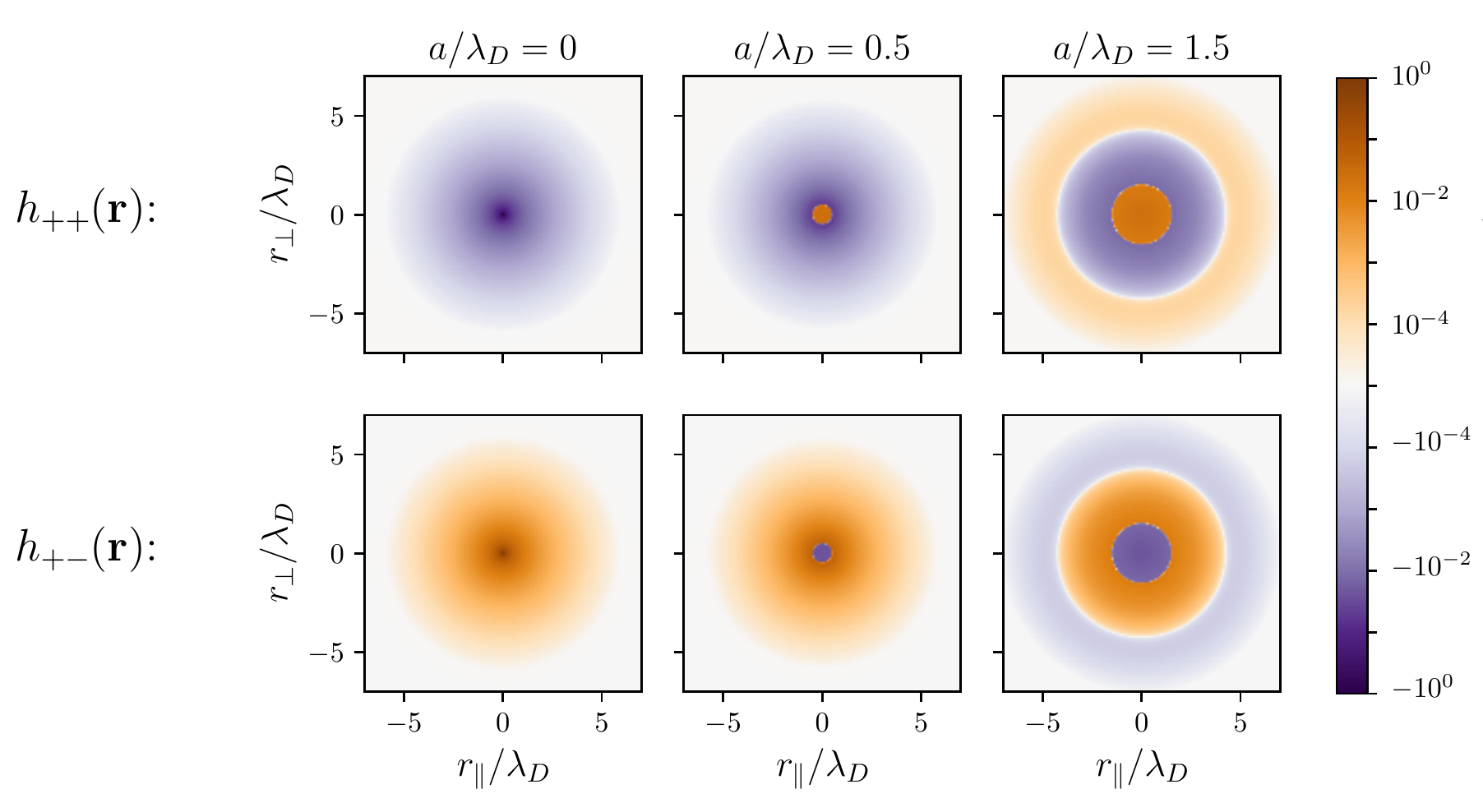} 
\caption{\textsf{Color plot of the pair-correlation functions, $h_{++}({\bf r})$ (top) and $h_{+-}({\bf r})$ (bottom), rescaled by $(2e^2z^2/k_B T \varepsilon_0 \varepsilon)^{3/2}\sqrt{n}$, for a symmetric ionic solution ($z_+=z_-=z$) at equilibrium. In cylindrical coordinates, the two axes, $r_{\parallel}/\lambda_D$ and $r_{\perp}/\lambda_D$, denote the axial position and radial distance, rescaled by the screening length, $\lambda_{D}$. The three columns differ by the value of $a/\lambda_D$, where $a$ is the finite ion-size parameter, as is indicated above each column.}}
\label{Fig2}
\end{figure}

\begin{figure}
\includegraphics[width = 0.77 \columnwidth,draft=false]{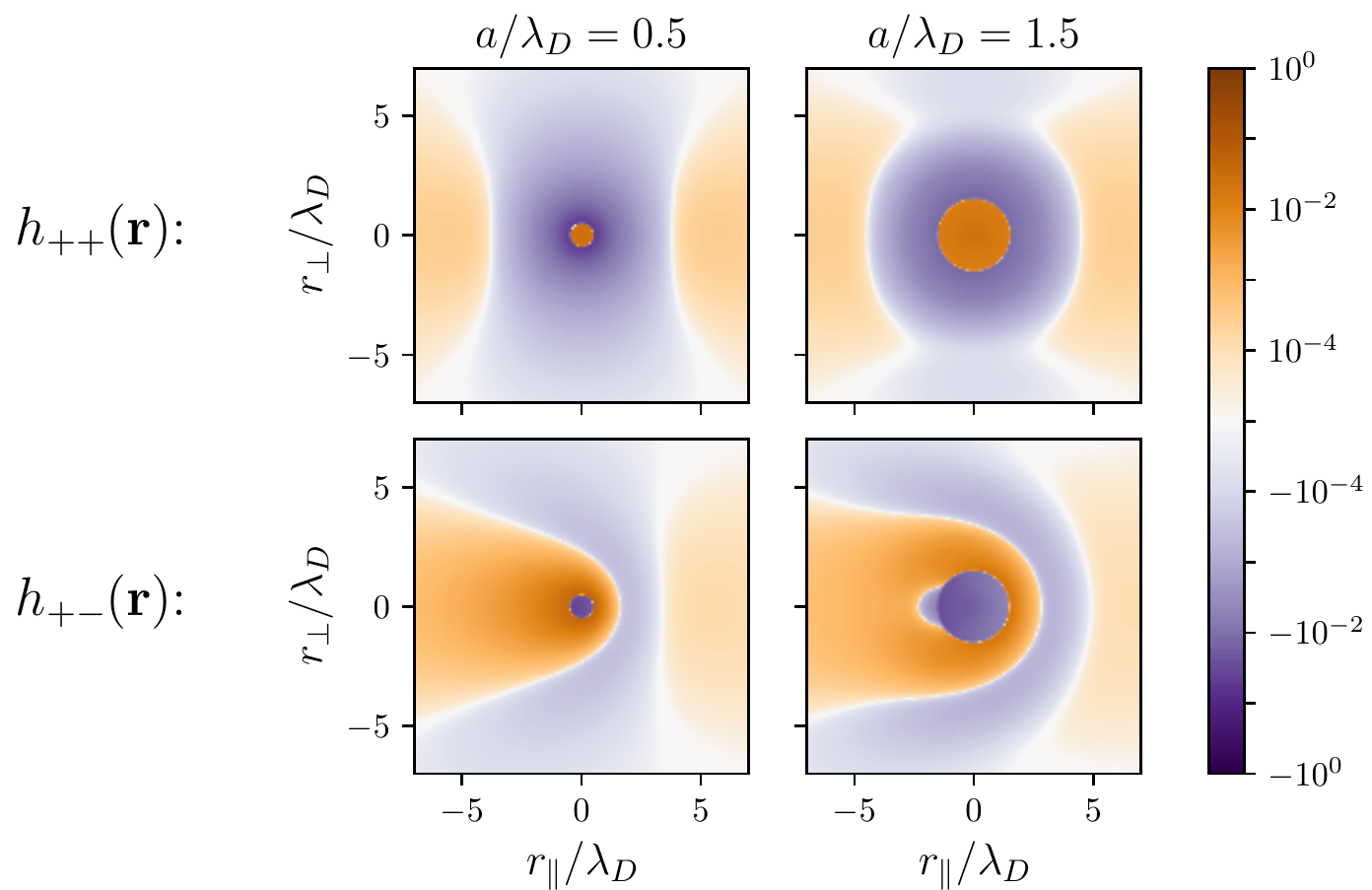} 
\caption{\textsf{The pair-correlation functions, $h_{++}({\bf r})$ and $h_{+-}({\bf r})$, as in Fig.~\ref{Fig2}, but driven out of equilibrium by an external electric field $E_0$ pointing in the $r_{\parallel}$ direction with rescaled field intensity $z \lambda_D/ l_E= 2$ (${E_0 =2k_B T/e z  \lambda_D}$). Left column: $a/\lambda_D= 0.5$; right column: $a/\lambda_D=1.5$.}}
\label{Fig3}
\end{figure}

A visualization of the correlations can be obtained by plotting the {\it pair-correlation function},
\beq \label{pair_corr}
h_{\alpha\beta}({\bf r})=\frac{1}{n_{\alpha}^{0}n_{\beta}^{0}}\langle\delta n_{\alpha}\left(0\right)\delta n_{\beta}\left({\bf r}\right)\rangle-\frac{\delta_{\alpha\beta}\delta\left({\bf r}\right)}{n_{\alpha}^{0}}=\frac{1}{\left(2\pi\right)^{3}}\int{\rm d}^{3}k\,\frac{\widehat{C}_{\alpha\beta}\left({\bf k}\right)}{n_{\alpha}^{0}n_{\beta}^{0}}{\rm e}^{-i{\bf k}\cdot{\bf r}},\eeq
where $\delta n_{\alpha}({\boldsymbol r},t)= n_{\alpha}({\boldsymbol r},t) -n^0_{\alpha}$. The pair-correlation function for symmetric ions is shown in Figs.~\ref{Fig2} and~\ref{Fig3}.
The correlation function is rescaled by $(2e^2/k_B T \varepsilon_0 \varepsilon)^{3/2}\sqrt{n}$ and depends on two parameters: $z \lambda_D/l_E$, that is the normalized electric field $E_0$, and $a/\lambda_D$, a rescaled cutoff length.

At equilibrium, $z \lambda_D/l_E=0$, and $h_{\alpha \beta}({\bf r})$ is spherically symmetric (Fig.~\ref{Fig2}). For $a/\lambda_D=0$ (point-like ion), the standard ionic atmosphere, $h_{\alpha \beta}(r) \propto {\rm e}^{-r /\lambda_D}/r$, is obtained. Equal charges are depleted, opposite charges are more abundant around the test charge, and the correlation function diverges at $r \to 0$. For small non-zero $a/\lambda_D$, the pair-correlation function behaves similarly to the point-like case, except that it has a finite value at $r \to 0$. The value is positive for $h_{++}(r)$ and $h_{--}(r)$ and negative for $h_{+-}(r)$. For larger $a/\lambda_D$ values ($1 \lesssim a/\lambda_D\lesssim 2.8$ for symmetric ions), the pair-correlation function decays in an oscillatory manner~(for a full derivation, see Ref.~\cite{Adar2019}). Similar damped oscillations were shown to exist at highly concentrated solutions~\cite{Mezger2008,Bazant2011,Kornyshev2008a}. When $a/\lambda_D$ is very large ($\gtrsim 2.8$ for symmetric ions, not shown in Fig.~\ref{Fig2}), the correlation function diverges with pure oscillatory modes leading to unphysical long-range order~\cite{Adar2019}. The divergence at a finite $a/\lambda_D$ value is a consequence of using the modified cutoff potential, Eq.~(\ref{u_co}). However, such high $a/\lambda_D$ values, are reached only at very high concentrations (as high as $9$\,M for NaCl in water at room temperature, which goes beyond crystallization), for which the use of the modified cutoff potential is unjustified (see appendix A).

When an electric field is applied to the system (Fig.~\ref{Fig3}), the pair-correlation function maintains rotational symmetry around the direction of the electric field, $\hat{E}_0$. Whereas $h_{++}(r)$ is symmetric under reflection with respect to the electric-field direction, the symmetry is broken for $h_{+-}(r)$. An ion moving in the direction of the electric field is likely to have an oppositely charged ion behind it, yet far enough from the excluded-volume region. On the other hand, there is a depletion area of oppositely charged ions in front of the moving ion at short distances, and at larger distances, oppositely charged ions are more abundant. For large $a/\lambda_D$ values, the electric field destroys the concentric rings of positive/negative charge density of the equilibrium pair-correlation function. Further analysis of the pair-correlation function in the presence of an applied field can be found in Ref.~\cite{Frusawa2022}.

Substituting Eq.~(\ref{C12}) in Eq.~(\ref{cond3}), and performing the angular part of the $k$-space integral, we obtain the following expressions for the conductivity corrections $\kappa_{{\rm hyd}}$ and $\kappa_{{\rm el}}$,
\beqa \label{kappa_el}
\kappa_{{\rm hyd}} =&&\frac{2\kappa_{0}}{\pi\gamma}\frac{r_{s}l_{E}^{3}}{\lambda_{D}^{3}}\int\limits _{-\infty}^{\infty}{\rm d}k\,\frac{g\left(k\right)\cos\left(ka\right)}{f^{2}\left(k\right)}\bigg\{3\gamma^{2}\sqrt{\frac{g\left(k\right)}{f\left(k\right)}}\left(1+\frac{\lambda_{D}^{2}}{l_{E}^{2}}\frac{f\left(k\right)}{g\left(k\right)}\right)\tan^{-1}\left(\frac{\lambda_{D}}{l_{E}}\sqrt{\frac{f\left(k\right)}{g\left(k\right)}}\right)\nonumber\\
 && \times\left(z_{+}z_{-}\cos\left(ka\right)+k^{2}\lambda_{D}^{2}\left(z_{+}^{2}+z_{-}^{2}\right)-\frac{f\left(k\right)h^{2}\left(k\right)}{\gamma^{2}g\left(k\right)}\right)\\
 && +\frac{\lambda_{D}}{l_{E}}\left[\frac{3f\left(k\right)h^{2}\left(k\right)}{g\left(k\right)}-3\gamma^{2}\left(z_{+}z_{-}\cos\left(ka\right)+\left(z_{+}^{2}+z_{-}^{2}\right)k^{2}\lambda_{D}^{2}\right)\left(1+\frac{2f\left(k\right)}{3g\left(k\right)}\lambda_{D}^{2}l_{E}^{-2}\right)\right]\bigg\}\nonumber\\\nonumber\\[5pt] 
 \kappa_{{\rm el}}=&&-\frac{4\kappa_{0}}{\pi}\gamma z_{+}z_{-}l_{B}l_{E}^{2}\int\limits _{-\infty}^{\infty}{\rm d}k\,\frac{k^{2}\cos^{2}\left(ka\right)h\left(k\right)}{f\left(k\right)}\left[1-\frac{l_{E}}{\lambda_{D}}\sqrt{\frac{g\left(k\right)}{f\left(k\right)}}\tan^{-1}\left(\frac{\lambda_{D}}{l_{E}}\sqrt{\frac{f\left(k\right)}{g\left(k\right)}}\right)\right]\nonumber
\eeqa
where $r_s= 1/(6\pi\eta\bar{\mu})$ is a reduced Stokes radius with $\bar{\mu}=(\mu_++\mu_-)/2$ and $l_B = e^2/(4\pi \varepsilon_0 \varepsilon k_B T)$ is the Bjerrum length. We can see from Eq.~(\ref{kappa_el}) that the rescaled conductivities, $\kappa_{\rm hyd}/\kappa_0$ and $\kappa_{\rm el}/\kappa_0$ depend on the ratios between the length-scales: $\lambda_D$, $l_B$, $r_s$, $l_E$ and $a$, on the valencies $z_{\pm}$, and the asymmetry parameter $\gamma$. Equation~(\ref{kappa_el}) is the main result of this paper. In the next sections, we will explore different limits and cases.

\subsection{The conductivity in the weak $E_0$ limit}\label{vanishing}
The first case that we would like to examine is the limit $\lambda_D/l_E \to 0$, {\it i.e.}, $E_0\ll  k_B T/e \lambda_D$. As an example, for aqueous solutions at room temperature with monovalent ions, $\lambda_D/l_E\simeq 100 \,E_0{\rm [V/ \AA]}/\sqrt{n_{\rm salt}{\rm [M]}}$, which means that the $\lambda_D/l_E\to 0$ limit occurs when  $E_0 \ll 10^{-4}\,{\rm  V/ \AA }=1\,{\rm  V/ \mu m }$ for $n_{\rm salt}=1$\,mM and ${E_0 \ll10^{-2}\,{\rm  V/ \AA }=100\,{\rm  V/ \mu m }}$ for $n_{\rm salt}=1$\,M.
In this limit, Eq.~(\ref{kappa_el}) reduces to
\beqa \label{result1}
&&\kappa_{\rm hyd}/\kappa_0=-\frac{r_{s}}{\lambda_{D}}\frac{2}{\pi\gamma}\int\limits _{-\infty}^{\infty} {\rm d}x\,\frac{\cos(\frac{ax}{\lambda_{D}})}{\cos(\frac{ax}{\lambda_{D}})+x^{2}}\nonumber\\ \nonumber\\
&&\kappa_{\rm el}/\kappa_0=-\frac{l_{B}}{\lambda_{D}}\frac{z_{+}z_{-}\gamma}{3\pi} \int\limits _{-\infty}^{\infty} {\rm d}x\,\frac{x^{2}\cos^{2}(\frac{ax}{\lambda_{D}})}{\left(\cos(\frac{ax}{\lambda_{D}})+x^{2}\right)\left(\frac{1}{2}\gamma\cos(\frac{ax}{\lambda_{D}})+x^{2}\right)},
\eeqa
where we used the change of variables $x=\lambda_D k$. Although the integrals in Eq.~(\ref{result1}) cannot be performed analytically, they can be easily computed numerically.

\begin{figure}
\includegraphics[width = 0.42 \columnwidth,draft=false]{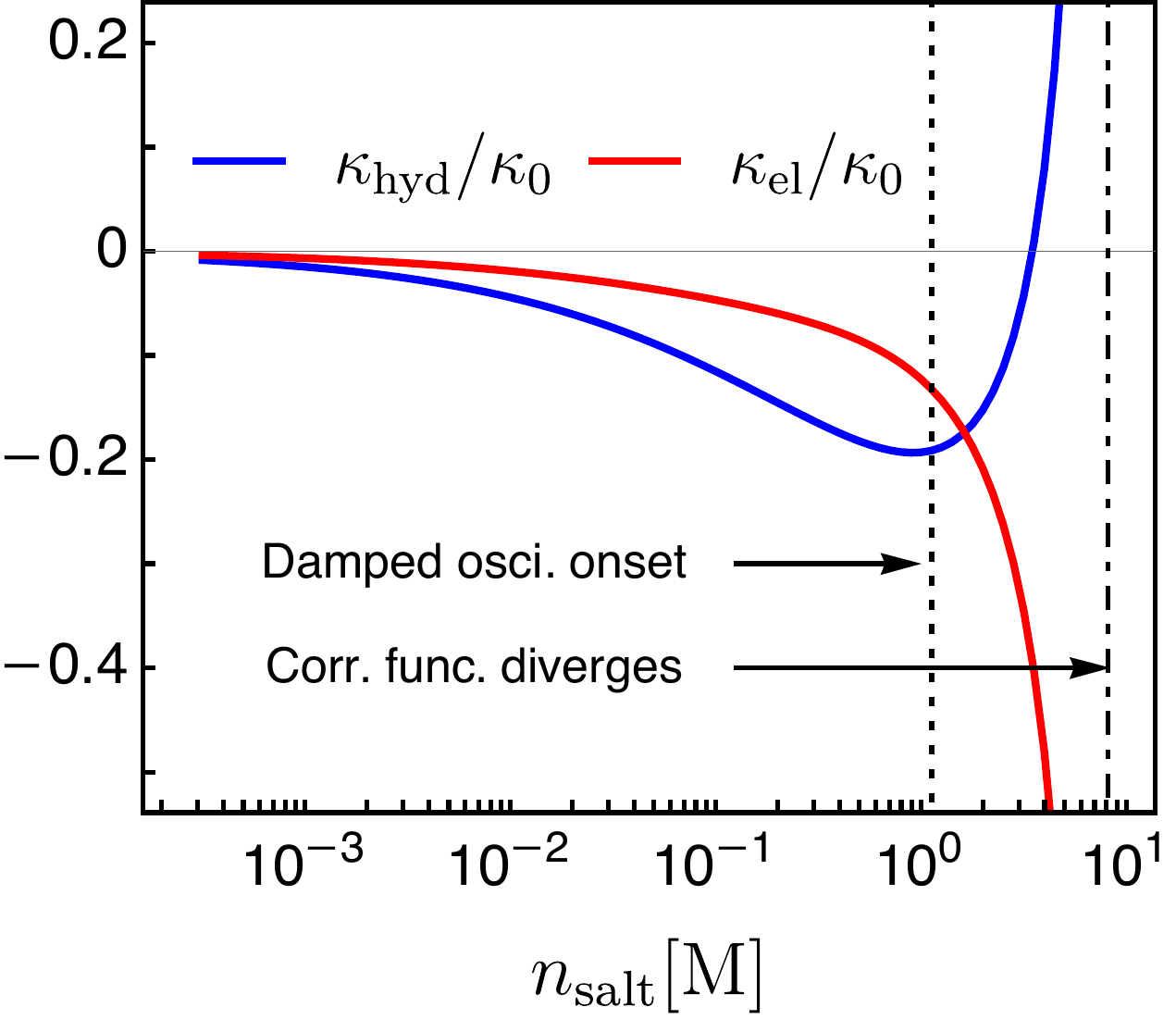} 
\caption{\textsf{The rescaled conductivity corrections, $\kappa_{\rm hyd}/\kappa_0$ (blue) and $\kappa_{\rm el}/\kappa_0$ (red) of monovalent salt solutions, as a function of the salt concentration $n_{\rm salt}$ on a semi-log plot. The conductivity corrections are calculated from Eq.~(\ref{result1}) with the parameters $l_B=7 {\rm \AA}$, $r_s=1.5 {\rm \AA}$ and $a=3 {\rm \AA}$. A vertical dotted line is plotted at the concentration where the correlation function displays damped oscillations and a vertical dotted-dashed line corresponds to the concentration where the correlation function, as well as $\kappa_{\rm hyd}/\kappa_0$ and $\kappa_{\rm el}/\kappa_0$, diverge.}}
\label{Fig4}
\end{figure}

The rescaled conductivity correction terms $\kappa_{\rm hyd}/\kappa_0$ and $\kappa_{\rm el}/\kappa_0$, are shown in Fig.~\ref{Fig4} on a semi-log plot as a function of $n_{\rm salt}$, the salt concentration, for monovalent salts, $z_{\pm}=1$. Both $\kappa_{\rm hyd}/\kappa_0$ and $\kappa_{\rm el}/\kappa_0$ approach zero in the infinite dilution limit ($n_{\rm salt}\to0$). One sees from the figure that $\kappa_{\rm hyd}/\kappa_0$ decreases as the concentration increases until a minimum is reached at $\sim 1\,{\rm M}$. Then, $\kappa_{\rm hyd}/\kappa_0$ increases until it diverges at a finite concentration. The minimum occurs very close to (but not exactly at) the onset of damped oscillations in the pair-correlation function, discussed earlier in Sec.~\ref{Two}, while the divergence occurs exactly when the correlation function diverges. The second correction term, $\kappa_{\rm el}/\kappa_0$, shows different behavior. It decreases as the salt concentration increases until it diverges to $-\infty$ at the same concentration where the correlation function diverges. We note that for $\gamma>2$, which is very uncommon for small inorganic ions, $\kappa_{\rm el}$ diverges prior to the diverges threshold of the correlation function, due to the term of $\frac{1}{2}\gamma\cos(ax/\lambda_{D})+x^{2}$ in the denominator of the $\kappa_{\rm el}/\kappa_0$ expression in Eq.~(\ref{result1}).

The two integrals of Eq.~(\ref{result1}) can be approximated for small $a$ by approximating $\cos(a x/\lambda_{D})\approx 1$ in their denominators. The integrals can then be calculated analytically using the residue theorem, yielding
\beq \label{approx}
\kappa/\kappa_0=1-\frac{r_{\rm s}}{\gamma \lambda_{\rm D}}{\rm e}^{-a/\lambda_{\rm D}}\,-\,\frac{z_{+}z_{-}\gamma}{12(1-\gamma/2)}\frac{ l_{B}}{\lambda_{\rm D}}\left(1-\sqrt{\frac{\gamma}{2}}+{\rm e}^{-2a/\lambda_{{\rm D}}}-\sqrt{\frac{\gamma}{2}}{\rm e}^{-\sqrt{2\gamma}a/\lambda_{{\rm D}}}\right).
\eeq
The divergence that occurs in the exact result at high concentrations is not present in the analytical approximation. By taking $a\to0$ in Eq.~(\ref{approx}), the DHO result for the conductivity is exactly recovered~\cite{OnsagerFuoss1932},
\beqa \label{DHO}
\kappa/\kappa_0=1-\frac{r_{{\rm s}}}{\gamma\lambda_{{\rm D}}}-\frac{z_{+}z_{-}\gamma}{6(1+\sqrt{\gamma/2})}\frac{l_{B}}{\lambda_{\rm D}},
\eeqa
where both $\kappa_{\rm hyd}/\kappa_0$ and $\kappa_{\rm el}/\kappa_0$ (second and third terms on the right-hand-side, respectively) are inversely proportional to $\lambda_D$.

\begin{figure}
\includegraphics[width = 0.45 \columnwidth,draft=false]{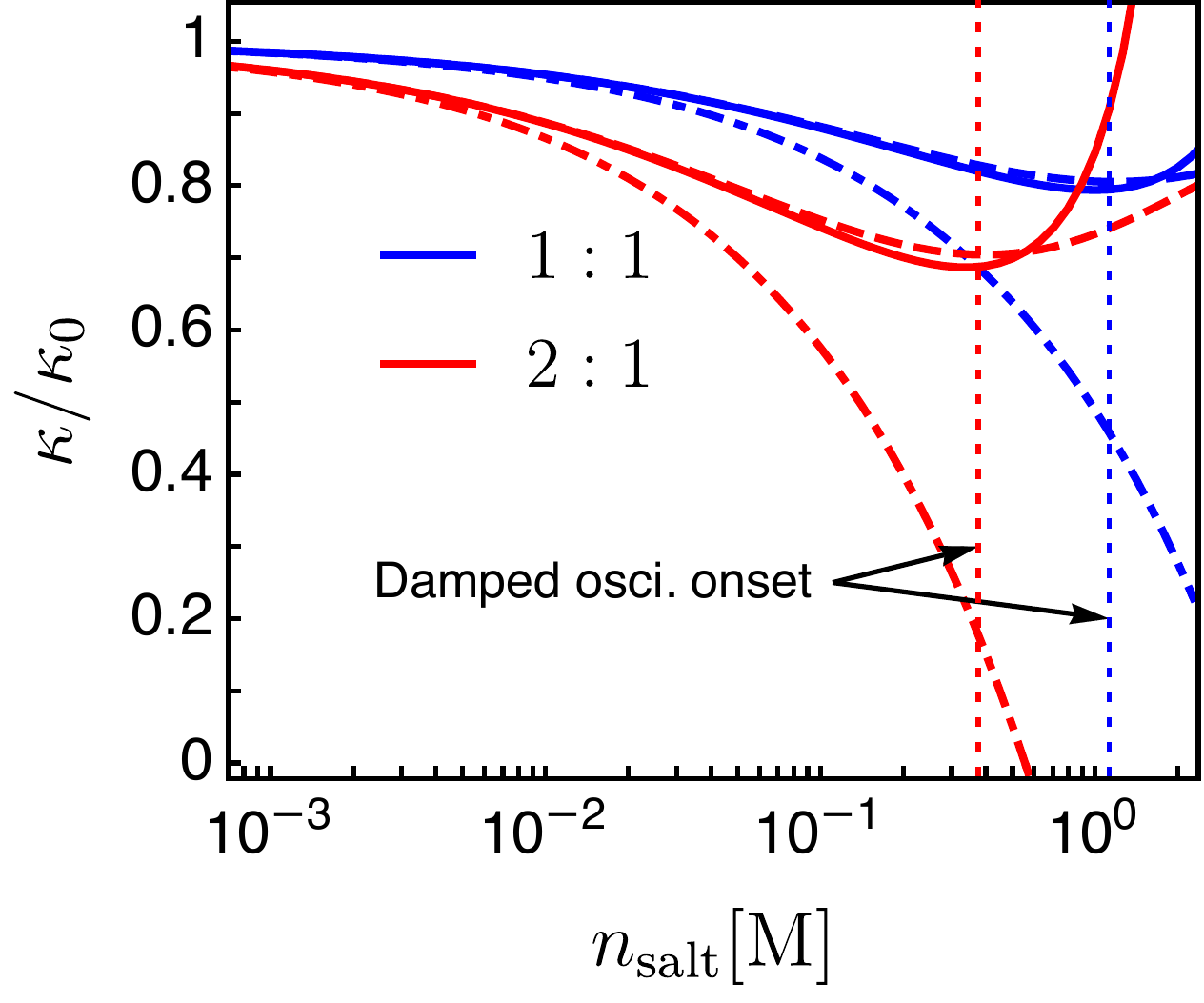} 
\caption{\textsf{The rescaled conductivity of 1:1 (blue) and 2:1 (red) electrolytes, as a function of the salt concentration $n_{\rm salt}=n^0_+$. Solid lines are numerical results, Eq.~(\ref{result1}), dashed lines are the analytic approximation, Eq.~(\ref{approx}) and dotted-dashed lines are DHO theory, Eq.~(\ref{DHO}). For the 1:1 case $\gamma=1$ and $r_s=1.5 {\rm \AA}$, while for the 2:1 case $\gamma=0.89$ and $r_s=2 {\rm \AA}$.
Other system parameters are: $l_B=7 {\rm \AA}$ and $a=3 {\rm \AA}$. Vertical dotted lines are plotted at the concentration where the correlation function displays damped oscillations for the 1:1 case (blue) and 2:1 case (red).}}
\label{Fig5}
\end{figure}

In Fig.~\ref{Fig5}, we explore the effect of multivalency on the conductivity by plotting the rescaled conductivity using our numerical results of Eq.~(\ref{result1}), the analytical approximation [Eq.~(\ref{approx})], and the DHO result [Eq.~(\ref{DHO})] of 1:1 monovalent electrolytes and 2:1 electrolytes ($+2e$ and $-e$ charges). We used a crude approximation that the mobility is inversely proportional to the valency (see Table~\Romannum{1} in Sec.~\ref{Comparison}) in order to estimate $\gamma$ and $r_s$, assigning a smaller $\gamma$ and larger $r_s$ for the 2:1 case. According to our numerical results, shown in Fig.~\ref{Fig5}, the rescaled conductivity decreases with multivalency, as is expected when the correlations become stronger. The threshold of decaying oscillations of the correlation function appears at lower concentrations for multivalent ions as compared with monovalent ions. The analytic approximation for the conductivity is shown to be in very good agreement with the numerical results below $1$\,M for the 1:1 case and below $\sim 0.5\,{\rm M}$ for the 2:1 case. Our results deviate substantially from the DHO result beyond a concentration of $\sim 10$\,mM, where a more pronounced deviation is seen for multivalent ions.

\subsection{The conductivity at finite electric fields for symmetric electrolytes} \label{finite_field}
\begin{figure}
\includegraphics[width = 0.46 \columnwidth,draft=false]{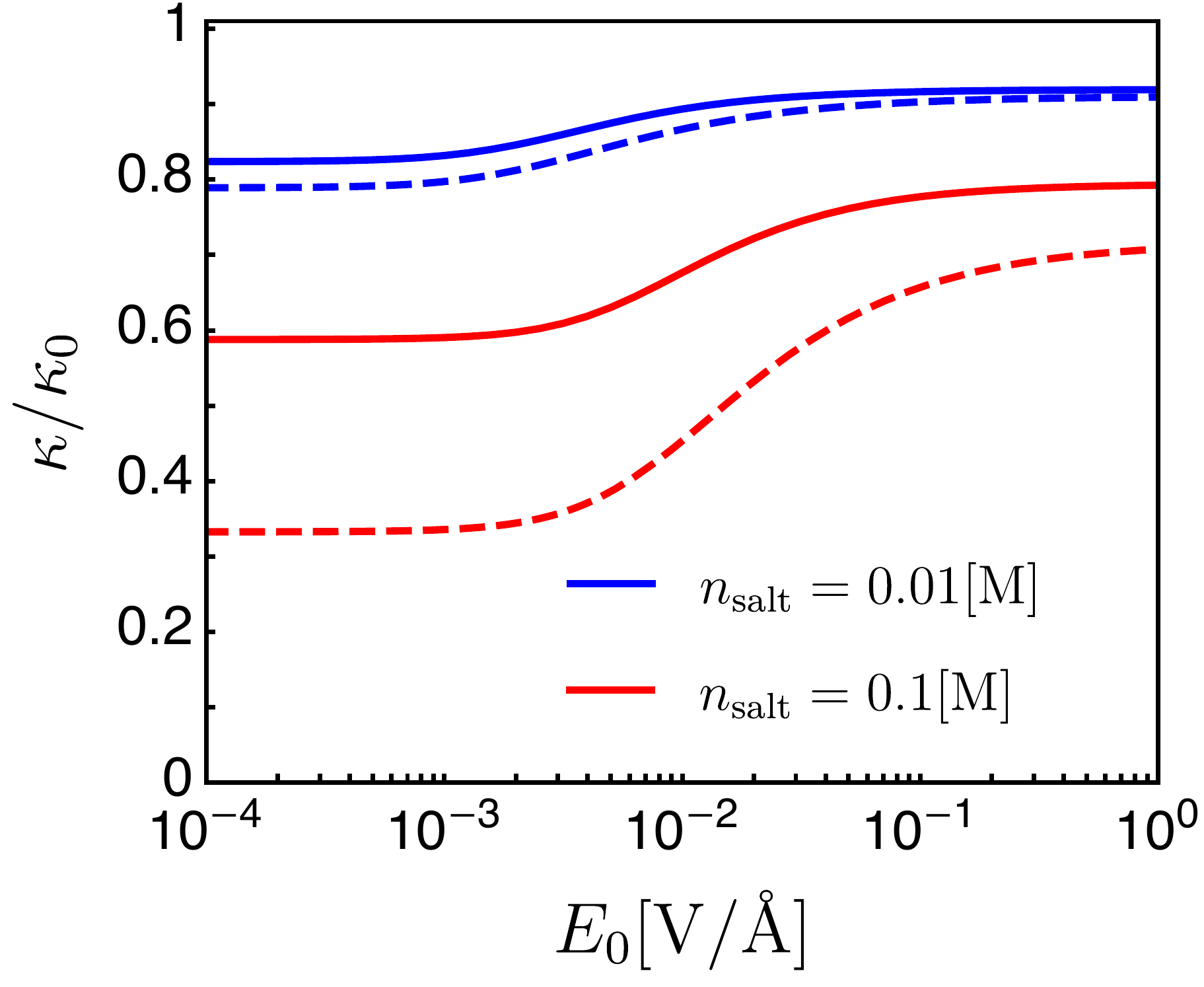} 
\caption{\textsf{(a) The rescaled conductivity $\kappa/\kappa_0$ of monovalent electrolytes as a function of the electric field intensity, plotted for two concentrations: $n_{\rm salt}=0.01\,$M in blue and $n_{\rm salt}=0.1\,$M in red. The system parameters are: $l_B = 1.7\,{\rm nm}$ (appropriate for methanol $\varepsilon_{\rm methanol}=32.7$ 
at room temperature), $r_s = 2.5\,{\rm \AA}$ and $a = 3\,{\rm \AA}$. Full lines are numerical results of Eq.~(\ref{electric_field1}) and dashed lines are OW theory of Eq.~(\ref{ons_E1}).
}}
\label{Fig6}
\end{figure}

For non-zero values of $E_0$, we keep $\lambda_D/l_E$ finite in Eq.~(\ref{kappa_el}) and assume for simplicity that the ions are symmetric, {\it i.e.} $z_+=z_-=z$, leading from Eq.~(\ref{gamma}) to $\gamma=1$. The rescaled conductivity correction terms become,
\small
\beqa \label{electric_field1}
\kappa_{\rm hyd}/\kappa_0  =&&-\frac{2r_{{\rm s}}}{\pi\lambda_{{\rm D}}}\int\limits _{0}^{\infty}{\rm d}x\frac{\cos\left(\frac{ax}{\lambda_{{\rm D}}}\right)}{\cos\left(\frac{ax}{\lambda_{{\rm D}}}\right)+2x^{2}}\left[1-\frac{3x^{2}}{2\xi^{2}}+\frac{\frac{3}{2}x^{2}\left(\cos\left(\frac{ax}{\lambda_{{\rm D}}}\right)+x^{2}+\xi^{2}\right)}{\xi^{3}\sqrt{\cos\left(\frac{ax}{\lambda_{{\rm D}}}\right)+x^{2}}}\tan^{-1}\frac{\xi}{\sqrt{\cos\left(\frac{ax}{\lambda_{{\rm D}}}\right)+x^{2}}}\right]\nonumber\\ \nonumber\\ 
 \kappa_{\rm el}/\kappa_0= &&-\frac{l_{{\rm B}} z^2}{\pi\lambda_{{\rm D}}}\int\limits _{0}^{\infty}{\rm d}x\frac{x^{2}\cos^{2}\left(\frac{ax}{\lambda_{{\rm D}}}\right)}{\frac{1}{2}\cos\left(\frac{ax}{\lambda_{{\rm D}}}\right)+x^{2}}\left[\frac{1}{\xi^{2}}  -\frac{1}{\xi^{3}}\sqrt{\cos\left(\frac{ax}{\lambda_{{\rm D}}}\right)+x^{2}}\tan^{-1}\frac{\xi}{\sqrt{\cos\left(\frac{ax}{\lambda_{{\rm D}}}\right)+x^{2}}}\right],
\eeqa
\normalsize
where $\xi\equiv z \lambda_D/l_E\propto E_0$.
In the $a/\lambda_D\to 0$ limit, the integrals can be performed analytically. A convenient way to perform the integrals is to take the $a\to0$ limit already in the 3-dimensional integral expressions of Eq.~(\ref{binary_kappa}), and then perform the radial integration in $k$-space before the angular part. The Onsager-Wilson (OW) result is then recovered~\cite{Onsager1957},
\beqa \label{ons_E1}
\kappa_{\rm hyd}/\kappa_0=&&-\frac{r_{{\rm s}}}{8\lambda_{{\rm D}}\xi^{3}}\bigg[\left(4\sqrt{2}\xi^{3}-3\sqrt{1+\xi^{2}}+3\sqrt{2}\right)\xi\nonumber\\
 &&+\,6\xi^{2}\sinh^{-1}(\xi)-3\left(1+2\xi^{2}\right)\tan^{-1}\left(\sqrt{2}\xi\right)+3\left(1+2\xi^{2}\right)\tan^{-1}\left(\frac{\xi}{\sqrt{1+\xi^{2}}}\right)\bigg]\nonumber\\\nonumber\\
 \kappa_{\rm el}/\kappa_0 = && \frac{l_{{\rm B}}z^2}{4\xi^{3}\lambda_{{\rm D}}}\bigg[\xi\left(\sqrt{2}-\sqrt{1+\xi^{2}}\right)-\tan^{-1}(\sqrt{2}\xi)+\tan^{-1}\left(\frac{\xi}{\sqrt{1+\xi^{2}}}\right)\bigg].
\eeqa
 \normalsize
In Fig.~\ref{Fig6}, we show the rescaled conductivity, $\kappa/\kappa_0=1-\kappa_{\rm hyd}/\kappa_0-\kappa_{\rm el}/\kappa_0$, according to Eq.~(\ref{electric_field1}), as a function of $E_0$, and in comparison to the OW result, Eq.~(\ref{ons_E1}). Two monovalent electrolyte concentrations are calculated: $n_{\rm salt}=0.01\,{\rm M}$ and $n_{\rm salt}=0.1\,{\rm M}$. As the electric-field effect is more pronounced when the ionic interactions are stronger, we used system parameters that correspond to solvents with a low dielectric constant (compared with $\varepsilon_{\rm water}\simeq80$), such as methanol with $\varepsilon_{\rm methanol}\simeq33$. The figure shows that the conductivity increases when $E_0$ is increased. This is a manifestation of the Wien effect, where the electric field lowers the ionic correlations.

Additionally, $\kappa/\kappa_0$ saturates at high electric fields. Such high fields are often not accessible experimentally, as they introduce other effects such as Joule heating~\cite{Joule_heating}. The relative increase in $\kappa/\kappa_0$, induced by the electric field, is more pronounced at high concentrations as compared with low concentrations. Our results predict that the relative increase in $\kappa/\kappa_0$ is smaller than the increase predicted by the OW theory. This is due to the suppression of the electrostatic interactions at short distances, included in our theory. As in the low electric field case analyzed in Sec.~\ref{vanishing}, the difference between our results for $\kappa/\kappa_0$ as compared with the $a\to0$ case (OW theory) increases with the ion concentration. We conclude that the Wien effect becomes more pronounced as the ion concentration increases, but to a lesser extent than the OW theory prediction.

By expanding Eq.~(\ref{electric_field1}) in powers of $\xi=z \lambda_{D}/l_{E}$ at weak electric fields, we see that the conductivity grows quadratically with $\xi\propto E_0$,
\beq
\frac{\kappa(\xi)-\kappa(0)}{\kappa_0}=\frac{2}{5\pi\lambda_{{\rm D}}}\int\limits _{0}^{\infty}{\rm d}x\frac{x^{2}\left[r_{{\rm s}}+l_{B}z^2\cos\left(\frac{ax}{\lambda_{{\rm D}}}\right)\right]\cos\left(\frac{ax}{\lambda_{{\rm D}}}\right)}{\left[x^{2}+\cos\left(\frac{ax}{\lambda_{{\rm D}}}\right)\right]^{2}\left[2x^{2}+\cos\left(\frac{ax}{\lambda_{{\rm D}}}\right)\right]}\xi^{2}
+\mathcal{O}(\xi^4).
\eeq
In the $\xi \to \infty$ limit, $\kappa_{\rm el}/\kappa_0\to 0$, while $\kappa_{\rm hyd}/\kappa_0$ approaches a constant value,
\beq
\lim_{\xi\to\infty}\kappa_{\rm hyd}/\kappa_0=-\frac{2r_{{\rm s}}}{\pi\lambda_{{\rm D}}}\int\limits _{0}^{\infty}{\rm d}x\frac{\cos\left(\frac{ax}{\lambda_{{\rm D}}}\right)}{\cos\left(\frac{ax}{\lambda_{{\rm D}}}\right)+2x^{2}}.
\eeq

The system behavior at strong electric fields can be understood from the correlation matrix, Eq.~(\ref{C_symmetric}). Taking the $\xi \to \infty$ limit is equivalent to $l_E\to 0$, and yields $\widehat{C}_{+-}(k)\to0$ and $\widehat{C}_{++}(k),\widehat{C}_{--}(k)\to-n\cos\left(ka\right)/[\cos\left(ka\right)+2k^{2}\lambda_{D}^{2}]$. The external electric field ``tears apart" pairs of oppositely charged ions and the correlation between such pairs vanishes in the $\xi \to \infty$ limit. However, since $E_0$ drags equally charged ions in the same direction, it does not destroy their (anti-) correlations. Thus, $C_{++}$ and $C_{--}$ remain finite. Since $\kappa_{\rm el}/\kappa_0$ is proportional to $C_{+-}$, it vanishes in the $\xi \to \infty$ limit. However, $\kappa_{\rm hyd}/\kappa_0$ depends on the difference between $C_{++}$ and $C_{+-}$. Therefore, it reaches a constant value in this limit.

We take this opportunity to clarify the relation of our results to two sub-cases of the Wien effect. In general, the Wien effect describes the increase in the conductivity with an increased electric field. Historically, it was addressed in two limiting cases: strong electrolytes (fully dissociated) at low concentrations, and weak electrolytes (mostly associated) at low {\it free-ion} concentrations. In the former case, which is sometimes called ``the first Wien effect" or simply ``the Wien effect", the increase is due to the destruction of the classical ionic cloud, which is independent of the ion size~\cite{Onsager1957}. This limiting case produces the OW result, Eq.~(\ref{ons_E1}). In the latter case, known as ``the second Wien effect", the increased conductivity is attributed to a modification in the dissociation kinetics that enhances the number of free ions~\cite{Onsager1934}. Onsager developed a theory for this second Wien effect, relying on Bjerrum's observation that when the concentration of free ions is small, the kinetics can be described by the law of mass action, regardless of the details of the interaction at short distances.

Our theory aims to describe a range of electrolytes and concentration regimes in which the ions are dissociated but can be significantly correlated. However, since our approach relies on a perturbative expansion, formally valid for {\it weak ionic correlations}, it is more accurate to regard it as the first Wien effect. The second Wien effect cannot be accounted for using the same perturbative expansion as it occurs in the opposite limit of very strong correlations (``ionic pairs").

\section{Comparison with experiments and simulations}\label{Comparison}
\begin{figure}
\includegraphics[width = 1 \columnwidth,draft=false]{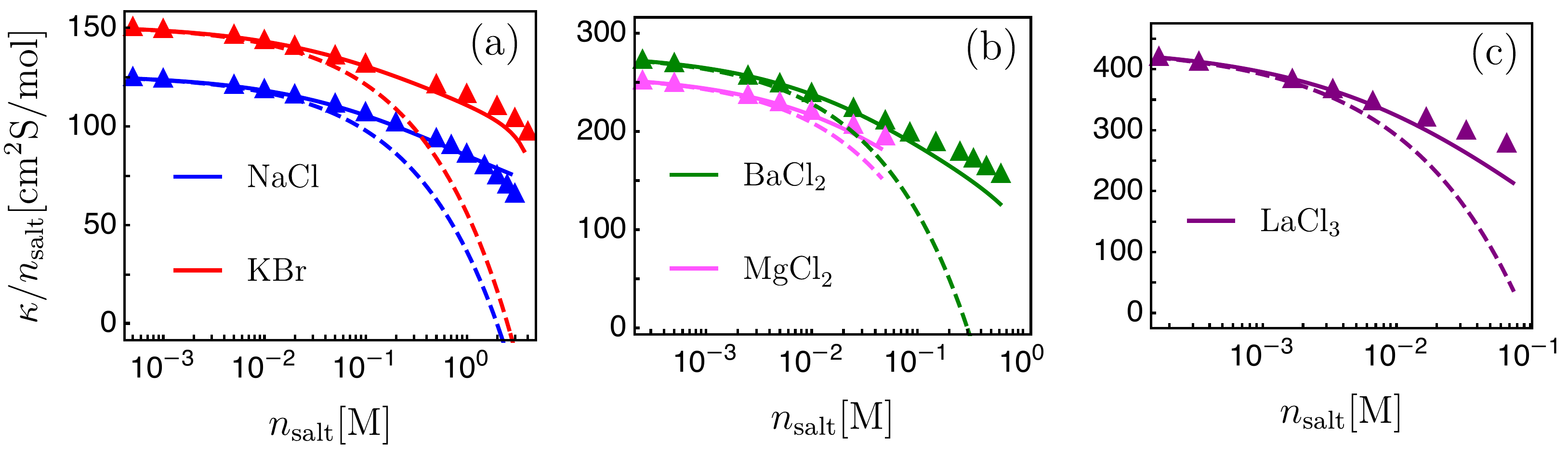} 
\caption{\textsf{The molar conductivity, $\kappa/n_{\rm salt}$, as a function of the salt concentration $n_{\rm salt}$. Two types of $1{:}1$ electrolytes are shown in (a); two types of $2{:}1$ electrolytes in (b); and one $3{:}1$ electrolyte in (c). Triangles are experimental data~\cite{Lide_Book,Lobo}; full lines are our numerical results, Eq.~(\ref{result1}); and dashed lines are the results obtained from DHO theory, Eq.~(\ref{DHO}). The electrolytes physical parameters are specified in Sec.~\ref{Comparison}.}}
\label{Fig7}
\end{figure}

In Fig.~\ref{Fig7}, we compare our numerical results for weak electric fields (Sec.~\ref{vanishing}) to experimental data for different aqueous ionic solutions at $T=25^\circ{\rm C}$, taken from Refs.~\cite{Lide_Book} and~\cite{Lobo}, where an extensive body of measurements is summarized. The {\it molar conductivity}, $\kappa/n_{\rm salt}$, which is commonly used in experiments, is plotted as a function of $n_{\rm salt}$ for 1:1 electrolytes (NaCl and KBr), 2:1 electrolytes (BaCl$_2$ and MgCl$_2$), and 3:1 electrolyte (LaCl$_3$). Note that the molar conductivity is proportional to the rescaled conductivity, $\kappa/\kappa_0$, since $\kappa_0$ is linear in $n_{\rm salt}$.
The electrolytes we consider have monovalent anions, $z_-=1$, and different cationic valencies $z_+$. Therefore, $n_{\rm salt}$ is the concentration of the cations, $n^0_{+}$.
At $T=25^\circ{\rm C}$, water viscosity is $\eta = 0.890\,{\rm mPa\cdot s}$~\cite{Korson1969} and the dielectric permittivity is $\varepsilon=78.3$~\cite{Malmberg1956}, yielding a Bjerrum length of $l_B=e^2/(4\pi \varepsilon_0 \varepsilon k_B T)=7.15\,\rm{\AA}$, and a screening length of $\lambda_{D}=1/\sqrt{4\pi l_{B}(z_+^2 n^0_++z_-^2 n^0_-)}=4.30\,[{\rm \AA}] /\sqrt{z_+(z_++1) n_{\rm salt}\,[{\rm M}]}$. In Table~\Romannum{1}, we summarize the values of the radii and diffusion coefficients at infinite dilution for all the ions considered in Figs.~\ref{Fig7}. In Table~\Romannum{2}, we present the electrolytes asymmetry parameter $\gamma$, cutoff length, a reduced Stokes radius $r_s$, and molar conductivity $\kappa_0/n_{\rm salt}$ at infinite dilution. They are all calculated from the parameters in Table~\Romannum{1} and the solution parameters $T$, $\varepsilon$, and $\eta$ mentioned above.
\begin{table}[h] \label{tab1}
\caption{The ion radii~\cite{Shannon1976} and diffusion coefficients for aqueous solutions at $T=25^\circ $C at infinite dilution limit~\cite{Lide_Book}. We use the ``Effective ionic radii" by Shannon with six-coordinate. Other sets for the ionic radii give very similar results~\cite{Shannon1976}.}

\begin{tabular}{ c  c  c }
 \hline\hline
 \,\,\,\,\,\,\,Ion\,\,\,\,\,\,\,\,\,&  \,\,\,\,\,\,$r \rm{[\AA]}$\,\,\,\,\,\,  &  \,\,\,\,\,\,$D  [10^{-5} \,{\rm cm}^2\, {\rm s}^{-1}]$\,\,\,\,\,\,  \\ [0.5ex]
 \hline
 Na$^{+}$ & $1.02$ & 1.334\\
 
 K$^{+}$ & $1.38$ & 1.957 \\
 
 Ba$^{2+}$ & $1.35$ & 0.847  \\
  
 Mg$^{2+}$ & 0.72 & 0.706 \\
 
 La$^{3+}$ & 1.03 & 0.619\\

 Cl$^{-}$ & 1.81& 2.032\\
  Br$^{-}$ & 1.96& 2.080\\
 \hline\hline
\end{tabular}
\end{table}
\begin{table}[h] \label{tab1}
\caption{The electrolytes asymmetry parameter $\gamma$ (Eq.~(\ref{gamma})), cutoff length ${a=r_+ + r_-}$, reduced Stokes radius ${r_s=k_B T/3\pi \eta (D_+ + D_-)}$, and molar conductivity at infinite dilution $\kappa_0/n_{\rm salt}$ (Eq.~(\ref{kappa_0}) with $n_{\rm salt}=n^0_+$) where S is the Siemens electric conductance unit, calculated from the parameters in Table~\Romannum{1} for aqueous solutions at $T=25^\circ $C.}
\begin{tabular}{ c  c  c  c c}
 \hline\hline
 \,\,\,\,\,\,\,\,\,\,\,\,Salt\,\,\,\,\,\,\,\,\,\,\,\,\,& \,\,\,\,\,\,\,\,$\gamma$\,\,\,\,\,\,\,\, & \,\,\,\,\,\,\,\,$a[{\rm \AA}]$\,\,\,\,\,\,\,\, & \,\,\,\,\,\,\,\,$r_s[{\rm \AA}]$\,\,\,\,\,\,\,\, & \,\,\,\,\,\,\,\,$\kappa_0/n_{\rm salt} [\rm{cm^2 \cdot S\cdot mol^{-1}}]$\,\,\,\,\,\,\,\,\\ [0.5ex]
 \hline
 NaCl & $1$ & 2.83  & 1.46 & 126.3\\

 KBr & $1$ &  3.34&  1.22 & 151.4\\

 BaCl$_2$ & 0.86 & 3.16 &1.70& 279.6 \\

 MgCl$_2$ & 0.84 & 2.53 &  1.79& 258.4\\

 LaCl$_3$ & 0.73 & 2.84 &  1.85& 437.6\\
\hline\hline
\end{tabular}
\end{table}
%

Figure~\ref{Fig7} shows that our numerical results are in good agreement with the experimental data at high concentrations, without any fit parameters. Furthermore, also shown in the figure, our numerical results present a significant improvement as compared with the DHO theory. However, the results become less accurate for multivalent ions at high concentrations. For 1:1 electrolytes, deviations exceed $ 5 \%$ only at concentrations above $\sim 2\,$M. For 2:1 electrolytes, deviations of $5\%$ emerge at concentrations above $\sim 0.1\,$M, while for 3:1 electrolytes, such deviations occur already at much smaller concentrations above $\sim 0.02\,$M. The inaccuracy of our results for multivalent ions at high concentrations has several causes. First, multivalent ions introduce very strong electrostatic interactions that break the perturbative calculation. Second, the modified potential does not approximate well the Coulomb potential with hardcore for multivalent ions at high concentrations, as demonstrated in Appendix~\ref{Testing} [see Fig.~\ref{Fig10}(a)]. Finally, the high charge density orders the liquid around the ions. The ordering changes the dielectric constant $\varepsilon$ and viscosity $\eta$, and these extra factors are not accounted for in our theory.

\begin{figure}
\includegraphics[width = 0.44 \columnwidth,draft=false]{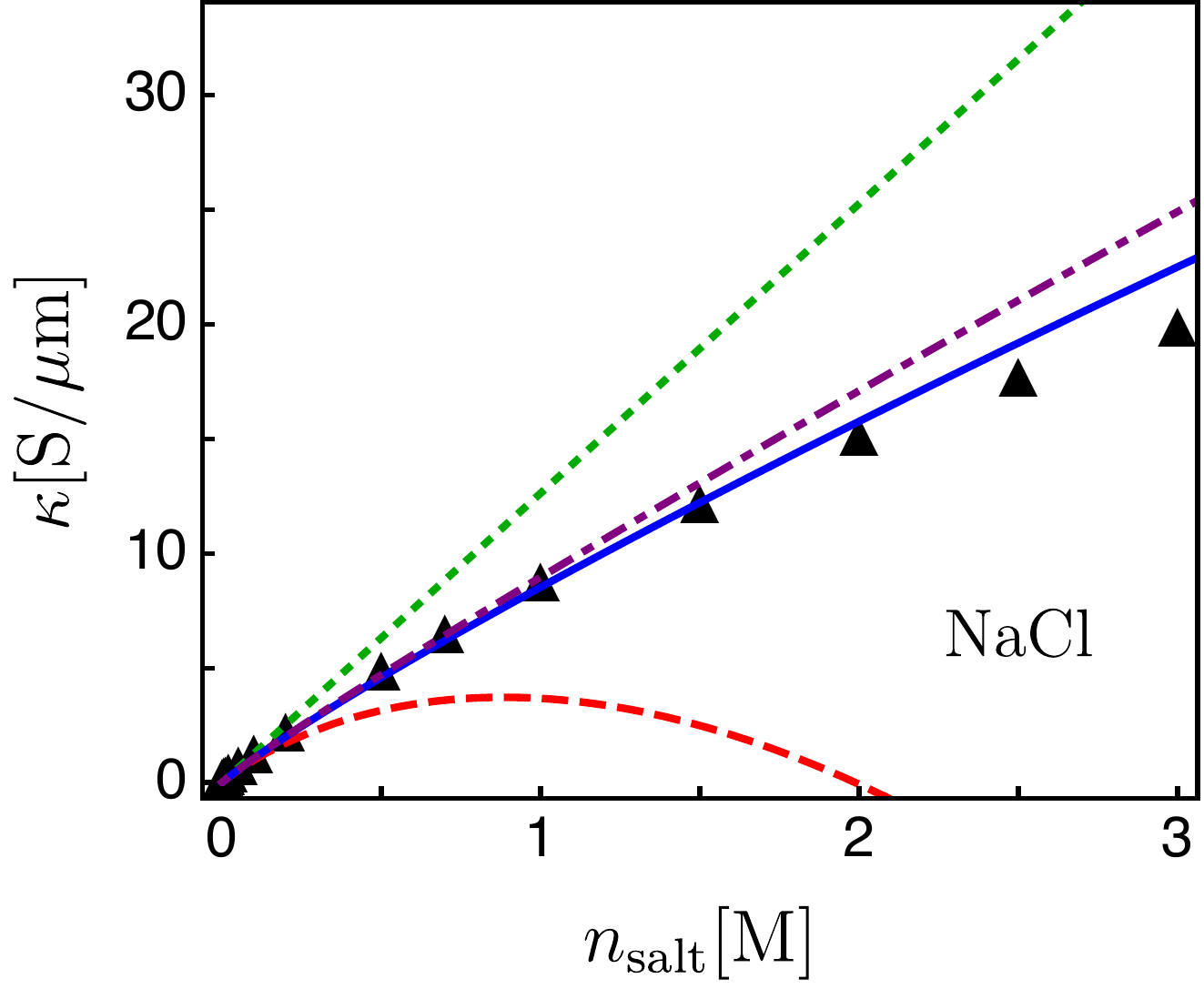}
\caption{\textsf{The conductivity, $\kappa$, of an aqueous solution of NaCl at $T=25^\circ{\rm C}$, as a function of the salt concentration $n_{\rm salt}$. Black triangles are the experimental data \cite{Lide_Book,Lobo}; green dotted line is the conductivity at infinite dilution, $\kappa_0$; full blue line is obtained numerically from Eq.~(\ref{result1}); dotted-dashed purple line is plotted from our analytical approximation, Eq.~(\ref{approx}); and dashed red line is obtained from DHO theory, Eq.~(\ref{DHO}). The electrolyte physical parameters are specified in Sec.~\ref{Comparison}}}
\label{Fig8}
\end{figure}

Figure~\ref{Fig8} summarizes different levels of approximation for the conductivity in units of $[{\rm S}/\mu {\rm m}]$ (rather than the rescaled conductivity). It shows experimental measurements of the conductivity of NaCl as a function of the concentration, compared with: (\romannum{1}) infinite dilution limit ($\kappa_0$) that is linear in $n_{\rm salt}$, (\romannum{2}) our numerical results [Eq.~(\ref{result1})], (\romannum{3}) our approximated results [Eq.~(\ref{approx})], and (\romannum{4}) the classical DHO theory [Eq.~(\ref{DHO})]. The numerical results are in excellent agreement with experimental measurements for concentrations up to $3\,$M. The analytical approximation also agrees quite well with the experimental data. As expected, the DHO theory and the infinite dilution limit deviate from the experimental measurements at high concentrations (substantial deviations occur above $\sim 0.5\,$M).

Our results in Sec.~\ref{finite_field} for the Wien effect at high concentrations should be compared with conductivity measurements at finite electric fields at high ionic concentrations. However, to the best of our knowledge, no experimental data are available in this regime. Moreover, little experimental data exist on the Wien effect even for dilute solutions. The reason, at least in part, is due to the experimental challenges involved in applying an external field while maintaining the system at a constant temperature.

\begin{figure}
\includegraphics[width = 0.47 \columnwidth,draft=false]{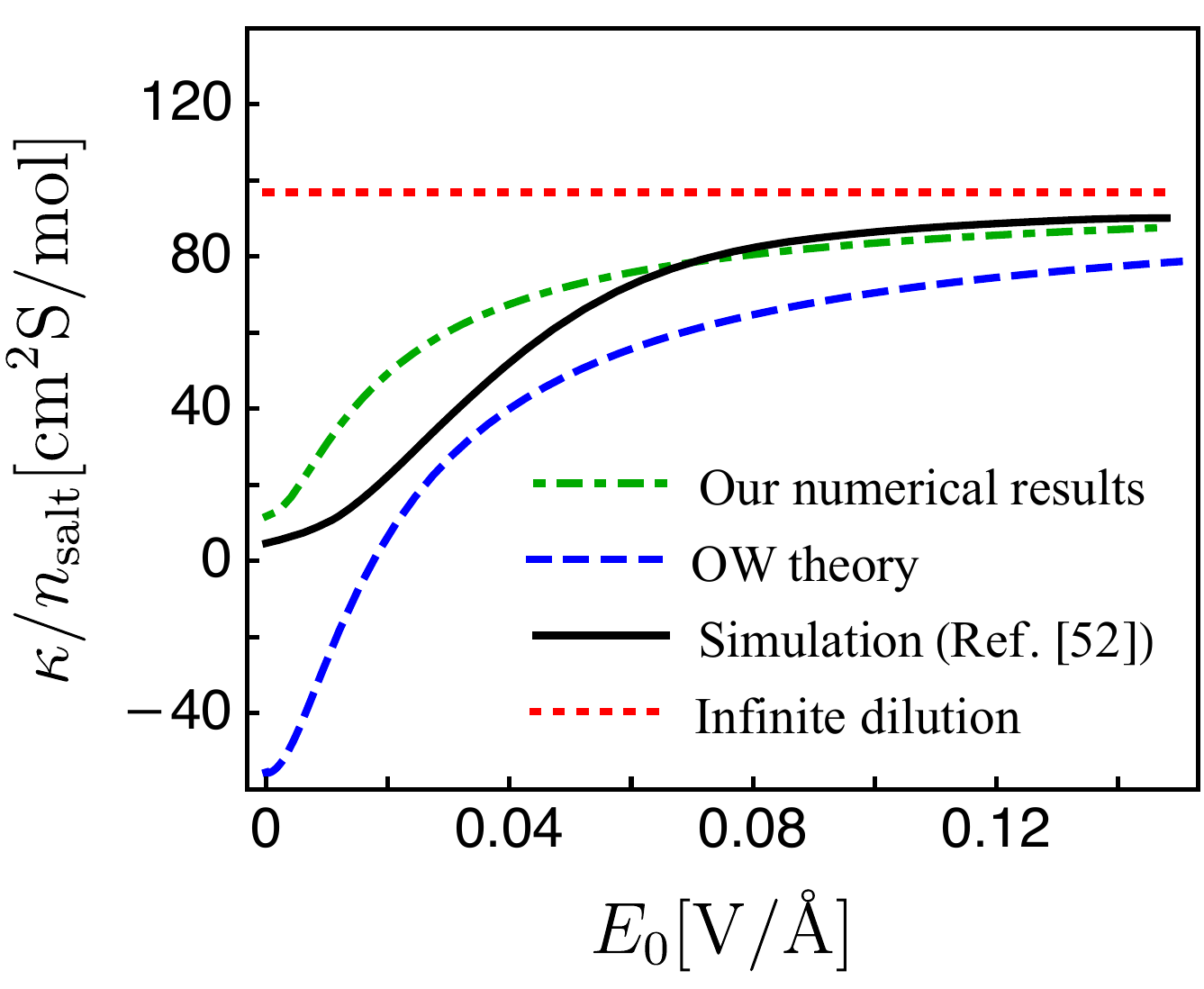} 
\caption{\textsf{The molar conductivity, $\kappa/n_{\rm salt}$ as a function of the electric field, $E_0$, excluding the hydrodynamic correction term, $\kappa_{\rm hyd}$ (in order to be consistent with the simulations, see Sec.~\ref{Comparison}). The system parameters match the implicit solvent simulation parameters in Ref.~\cite{Lesnicki2021}: $z_+=z_-=1$, $n_{\rm salt}=0.1\,$M, $T=300\,$K, $\varepsilon = 10$, $\bar{\mu} = 3.14\,{\rm s/Kg}$ and $a=r_++r_-=3.49\,{\rm \AA}$. The molar conductivity in our numerical results~(Eq.~(\ref{electric_field1})) and OW theory~(Eq.~(\ref{ons_E1})), are plotted without their respective hydrodynamic correction term, and compared with simulation data with implicit solvent, taken from Ref.~\cite{Lesnicki2021}, and to the infinite dilution limit.}}
\label{Fig9}
\end{figure}

Recently, field-dependent ionic conductivities were calculated from molecular dynamics simulations, using generalized fluctuation-dissipation relations~\cite{Lesnicki2020,Lesnicki2021}. This method yields the differential conductivity, ${\kappa_{\rm diff} \equiv {\rm d}\langle J_{\parallel}\rangle/{\rm d}E_0}$, related to the standard conductivity, ${\kappa=\langle J_{\parallel}\rangle/E_0}$, by ${\kappa=(1/E_{0})\int_{0}^{E_{0}}\kappa_{{\rm diff}}(E)\,{\rm d}E}$. In Fig.~\ref{Fig9}, the simulation results of Ref.~\cite{Lesnicki2021} for the molar conductivity are reproduced, where the differential conductivity is converted by integration to the standard conductivity. The simulations take into account the solvent only implicitly and do not account for the conductivity correction due to the counterflow of the solvent. Thus, they are compared with our numerical results, Eq.~(\ref{electric_field1}) and the OW theory, Eq.~(\ref{ons_E1}), {\it without} the hydrodynamic correction term, $\kappa_{\rm hyd}$. While our numerical results deviate significantly from the simulations, they describe the same qualitative behavior and are in much better agreement with the simulations than the OW theory is. In particular, the simulations support our prediction that at high ionic concentrations, the relative increase of the conductivity due to the Wien effect is smaller as compared with the increase predicted by the OW theory. We note that for system parameters as in Fig.~\ref{Fig9}, the OW result does not make sense as it predicts negative conductivity for weak electric fields.

The deviation of our results from the simulations might be attributed to the fact that the low dielectric constant used in the simulation, $\varepsilon=10$, induces very strong inter-ionic correlations in the electrolyte. This can be seen from the conductivity at zero electric field shown in Fig.~\ref{Fig9}, which is only a small percent of the fully uncorrelated case, represented by the infinite dilution limit. For such strong correlations, the perturbative expansion in Sec.~\ref{conductivity_calc}, and the use of the modified cutoff potential, are unjustified. It might be more appropriate to describe this low-$\varepsilon$ case via the second Wien effect (see discussion in Sec.~\ref{finite_field}), which is beyond the scope of this paper.

\section{Conclusions}\label{Conclusions}
This paper presents a theory for conductivity at high electrolyte concentrations, applicable for multivalent ions and finite electric fields. We used a stochastic density functional theory (SDFT) and a modified electrostatic potential that suppresses the unphysical short-range attraction between oppositely charged ions. At low electric fields, the theory is particularly accurate for monovalent salts, showing excellent agreement with experimental data at concentrations as high as a few molars with no fit parameters. Its range of applicability decreases for multivalent ions due to the strong electrostatic interactions that break the perturbative approach, and the inaccuracy of the modified potential at high concentrations. Nevertheless, the theory provides accurate predictions for 2:1 and 3:1 electrolytes up to concentrations of $\sim0.1$\,M and $\sim0.02$\,M, respectively, without any fit parameters. This is far beyond the applicability range of the well-known Debye-H\"uckel-Onsager (DHO) theory. 

For strong electric fields, we recover the Wien effect and show that similarly to dilute solutions, the conductivity at high ionic concentrations displays a sigmoid-like behavior, where the conductivity will make a transition between two limiting values when the field strength increases. The relative increase in the conductivity at high concentrations is smaller than the increase predicted by the WO theory due to the suppression of the electrostatic interactions at short distances. Recent simulations done in this concentrated regime with an implicit solvent show that our results present an improvement over the WO theory in capturing the Wien effect. In order to further test the theory, experiments for high concentrations at finite electric fields are needed.

Our results are obtained by linearizing the equations of motion, Eqs. (\ref{continuity})-(\ref{j}) and (\ref{Stokes0}), around the mean-field solution. In order to obtain more accurate results at high ionic concentrations, the equations of motion can be expanded systematically beyond linearization. The next order term can be formally evaluated, but its expression requires the analytical form of the fluctuating concentrations obtained from the linearized equations, which makes it very complicated.

In principle, one could solve the full equations of motion numerically at each time step by dividing space into small cells. However, as the cell size decreases, the noise becomes stronger and makes the concentrations $n_\alpha({\bf r},t)$ negative at certain iterations, which cannot be handled numerically due to the square root in Eq.~(\ref{j})~\cite{Russo2021}. Small cell size is necessary to numerically account for high ionic concentrations since important effects occur at small length scales (of order $a$ and $\lambda_D$). Therefore, a numerical solution at high ionic concentration cannot be obtained to the best of our knowledge.

An interesting extension to the work presented here would be to study the conductivity at high concentrations but in lower dimensions. For example, carrying out similar calculations in 1D or 2D can shed light on the dynamics in nanofluidic pores or slits, respectively~\cite{Kavokine2019, Robin2021}. The electrostatic potential in lower
dimensions has a very different behavior than in 3D. For example, in 1D it grows linearly with distance, which causes two ions to be permanently bound. In realistic pores and slits, there is a crossover from the 1D and 2D potential at short distances to the 3D one at long distances~\cite{Kavokine2021}. Such a potential can be included in $v_{\alpha \beta}(r)$ in Sec.~\ref{formalism} in a straightforward manner. In addition, the hydrodynamic flow in pores and slits is also different since it includes the hydrodynamic interaction with the walls. It remains to be seen what are the consequences of these differences and what possible analogs of the Wien effect would look like in systems of lower dimensionality.

\bigskip\bigskip
{\em Acknowledgments}~~
We would like to thank R. Adar, M. Bazant, L. Bocquet, H. Bonneau, V. D\'emery, A. Donev, A. Kornyshev, R. Netz, P. Robin,  G. Yossifon, and B. Zaltzman for correspondence and suggestions, and B. Rotenberg for discussing with us his simulation results. Y. A. is thankful for the support of the Clore Scholars Programme of the Clore Israel Foundation. This work was supported by the Israel Science Foundation (ISF) under Grant No. 213/19 and by the National Natural Science Foundation of China (NSFC) -- ISF joint program under Grant No. 3396/19.

\bigskip\bigskip
{\em Data Availability}~~
The data that support the findings of this study are available within the article.

\appendix

\section{Testing the modified potential}\label{Testing}
In our model, we use the modified potential in Eq.~(\ref{u_co}), instead of the Coulombic potential with a hardcore repulsion, Eq.~(\ref{hardcore}). The latter cannot be used as it breaks the perturbative approach (see Sec.~\ref{conductivity_calc}). In order to justify the use of the modified potential, we compare it to that of the Coulomb with hardcore potential, in a simplified system that can be solved without a perturbative calculation.

We consider two ions of charge $q_{1}$ and $q_{2}$ and radius $r_1$ and $r_2$, inside a sphere of radius $R$. The first ion is fixed at the center of the sphere while the second ion is free to move around. We calculate the average distance between the two ions from the relation,
\beq \label{average}
\langle r\rangle=\frac{\int\limits _{0}^{R}{\rm d}r\,r^{3}{\rm e}^{-v\left(r\right)/k_B T}}{\int\limits _{0}^{R}{\rm d}r\,r^{2}{\rm e}^{- v\left(r\right)/k_B T}}.
\eeq
For the Coulomb with hardcore interaction, we use the potential
\beq \label{hardcore_app}
v_{\rm c}^{\rm hc}(r)=\begin{cases}
\frac{q_{1}q_{2}}{4\pi\varepsilon_{0}\varepsilon r} & r>a\\
\infty & {\rm else}
\end{cases}
\eeq
where $a=r_1+r_2$, while for the modified Coulomb with a cutoff potential, we use
\beq \label{u_co_app}
v_{\rm co}\left(r\right)=\frac{{q_{1}q_{2}}}{4\pi\varepsilon_{0}\varepsilon r}\Theta\left(r-a\right),
\eeq
where $\Theta(r)$ is the Heaviside function.

Calculating $\langle r \rangle$ with $v_{\rm c}^{\rm hc}(r)$ and $v_{\rm co}(r)$ yields $\langle r\rangle_{\rm c}^{\rm hc}$ and $\langle r\rangle_{\rm co}$, respectively.
The normalized deviation,
\begin{equation} \label{sigma}
\sigma\equiv \frac{\langle r\rangle_{\rm co}-\langle r\rangle_{\rm c}^{\rm hc}}{\langle r\rangle_{\rm c}^{\rm hc}}
\end{equation}
quantifies how well the modified potential approximates the Coulombic potential with a hardcore repulsion, where $\sigma=0$ indicates perfect agreement between the two potentials, while large deviations from zero indicate poor agreement. The normalized deviation $\sigma$ depends on the sphere radius $R$. This radius can be related to a multi-ion electrolyte system by dividing the electrolyte into cells, where each cell (represented by our sphere) contains on average two ions. The relation between $R$ and the ionic concentrations is then
\beq
(4\pi/3)R^3(n^0_{+}+n^0_{-})\approx 2.
\eeq
For a $z_+{:}1$ electrolyte, $n_{\rm salt}=n^0_+=n^0_-/z_+$, and thus,
\beq \label{n_star}
n_{\rm salt}^{*}=\frac{3}{2\pi}\frac{1}{z_{+}+1}R^{-3}.
\eeq
The salt concentration, $n_{\rm salt}^*$, is an approximation of the electrolyte salt concentration of the real system, $n_{\rm salt}$.

\begin{figure}
\includegraphics[width = 0.92 \columnwidth,draft=false]{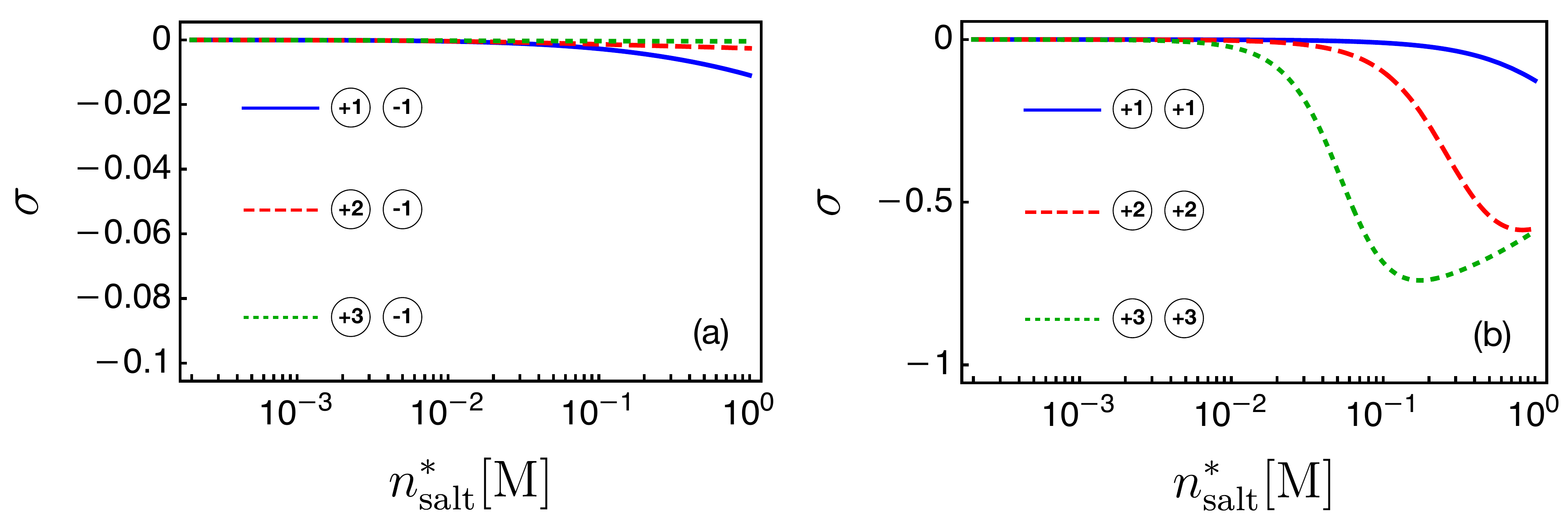} 
\caption{\textsf{The normalized deviation, $\sigma$ [Eq.~(\ref{sigma})], of the average distance between two ions interacting inside a sphere via the modified cutoff potential [Eq.~(\ref{u_co_app})], from the same distance but for ions that interact via the Coulomb with hardcore potential [Eq.~(\ref{hardcore_app})], as a function of $n_{\rm salt}^{*}$, defined in Eq.~(\ref{n_star}). The case of cation-anion is shown in (a), while the cation-cation case is shown in (b), for monovalent anions and different cationic valencies. The other parameters are $l_B=7\,{\rm \AA}$ and $a=3\,{\rm \AA}$.}}
\label{Fig10}
\end{figure}

Figure~\ref{Fig10} shows the normalized deviation $\sigma$ calculated from Eq.~(\ref{average}) as a function of $n_{\rm salt}^*$, with system parameters appropriate for simple salts in water at room temperature. Figure~\ref{Fig10}a shows the case of cation-anion interactions and Fig.~\ref{Fig10}(b) cation-cation interactions, for three electrolytes: 1:1, 2:1 and 3:1. For the cation-anion case, $\sigma$ barely deviates from $0$, even at $1$\,M concentration, indicating that the modified potential approximates well the Coulomb with hardcore interaction for opposite charges. For comparison, we note that a pure Coulomb interaction, $v_c(r)={q_{1}q_{2}}/4\pi\varepsilon_{0}\varepsilon r$, leads to $\langle r \rangle_{\rm c}=0$ and a normalized deviation of $(\langle r \rangle_{\rm c}-\langle r\rangle_{\rm c}^{\rm hc})/\langle r\rangle_{\rm c}^{\rm hc}=-1$ at any concentration (see supplementary material in Ref.~\cite{Avni2022}).

For cation-cation interactions, $\sigma$ is close to zero at intermediate concentrations, and starts to deviate substantially from zero at high concentrations, where the deviation increases with the cation valency. For a 1:1 electrolyte, $\sigma<-0.1$ for $n_{\rm salt}^*\gtrsim 0.8$, while for 2:1 and 3:1 electrolytes it occurs at $n_{\rm salt}^*\gtrsim0.1$ and $n_{\rm salt}^*\gtrsim0.02$, respectively. The failure of the modified potential to mimic the coulomb with hardcore potential at very high concentrations explains, at least in part, the inaccuracy of the theory in comparison to experimental data at high concentrations, as shown in Fig~\ref{Fig7}.

\section{Testing the single cutoff approximation}\label{different cutoffs}
We explore the validity of the simplification done in Sec.~\ref{Two}, where the species-dependent cutoff length, $r_{\alpha} + r_{\beta}$, is replaced by a single cutoff, $a=r_+ + r_-$, for any ionic pair of $\alpha$ and $\beta$. Without this simplification, the rescaled conductivity correction terms in the weak electric-field limit of a binary electrolyte (following the analysis of Sec.~\ref{model}) are

\small
\beqa \label{hyd3a}
 \kappa_{\rm hyd}/\kappa_0=-\frac{2r_{s}z_{+}z_{-}}{\pi\lambda_{D}\gamma}\int\limits _{0}^{\infty} &&\frac{x^{2}\left(\frac{z_{+}}{z_{-}}\cos\left(\frac{a_{+}x}{\lambda_{D}}\right)+\frac{z_{-}}{z_{+}}\cos\left(\frac{a_{-}x}{\lambda_{D}}\right)+2\cos\left(\frac{a_{+-}x}{\lambda_{D}}\right)\right)-\sin^{2}\left(\frac{\left(a_{+}-a_{-}\right)x}{2\lambda_{D}}\right)}{8x^{4}\bar{z}^{2}+4\bar{z}x^{2}\left(z_{+}\cos\left(\frac{a_{+}x}{\lambda_{D}}\right)+z_{-}\cos\left(\frac{a_{-}x}{\lambda_{D}}\right)\right)+z_{+}z_{-}\left(\cos\left(\frac{\left(a_{+}-a_{-}\right)x}{\lambda_{D}}\right)-1\right)}
\nonumber \\ \nonumber\\ \\
 \kappa_{\rm el}/\kappa_0=-\frac{l_{{\rm B}}\gamma}{3\pi\lambda_{{\rm D}}}\int\limits _{0}^{\infty} && {\rm d}x\frac{x^{4}z_{+}z_{-}\bar{z}^{2}\cos^{2}\left(\frac{a_{+-}x}{\lambda_{{\rm D}}}\right)}{\left(\bar{z}x^{2}+\frac{1}{2}z_{+}\cos\left(\frac{a_{+}x}{\lambda_{D}}\right)\right)\left(\bar{z}x^{2}+\frac{1}{2}z_{-}\cos\left(\frac{a_{-}x}{\lambda_{{\rm D}}}\right)\right)-\frac{1}{4}z_{-}z_{+}\cos^{2}\left(\frac{a_{+-}x}{\lambda_{{\rm D}}}\right)}\nonumber\\
 && \times\frac{1}{\left(\mu_{+}z_{+}\cos\left(\frac{a_{+}x}{\lambda_{{\rm D}}}\right)+\mu_{-}z_{-}\cos\left(\frac{a_{-}x}{\lambda_{{\rm D}}}\right)\right)/\left(8\bar{\mu}\bar{z} \right)+\frac{1}{2}x^{2}}\,,\nonumber
\eeqa
\normalsize
where we defined $a_{+}=2r_+$, $a_{-}=2r_-$ and $a_{+-}=r_++r_-$.

\begin{figure}
\includegraphics[width = 1 \columnwidth,draft=false]{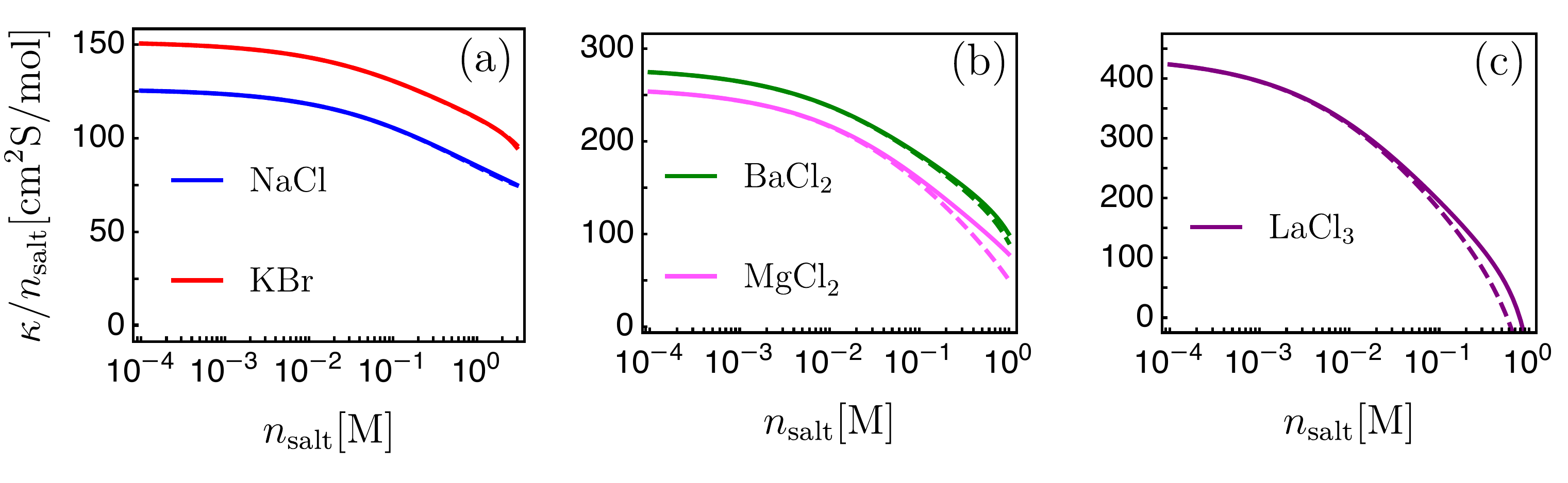} 
\caption{\textsf{The molar conductivity, $\kappa/n_{\rm salt}$, of $1{:}1$ (a), $2{:}1$ (b) and $3{:}1$ (c) electrolytes (the same electrolytes as in Fig.~\ref{Fig7}), as a function of the salt concentration $n_{\rm salt}$, shown for the single cutoff length approximation [full lines, Eqs.~(\ref{result1})] and when using three different cutoffs [dashed lines, Eq.~(\ref{hyd3a})]. The electrolyte physical parameters are specified in Sec.~\ref{Comparison}}.}
\label{Fig11}
\end{figure}

In Fig.~\ref{Fig11}, we show the molar conductivity, $\kappa/n_{\rm salt}$, for the same five electrolytes analyzed in Fig.~\ref{Fig7}, and compare the results of using three different cutoffs [Eq.~(\ref{hyd3a})], to the single cutoff length case $a=a_{\pm}=a_{+-}$ [Eq.~(\ref{result1})]. For concentrations below $\sim 0.05\,$M, the difference is negligible for all considered electrolytes. Above $\sim0.05\,$M, there is a small yet visible difference, but only for electrolytes with trivalent ions (LaCl$_3$) or electrolytes with large size asymmetry between the cation and anion (MgCl$_2$, where $r_-\approx2.5r_+$, see Table~\Romannum{1} in Sec.~\ref{Comparison}). In these cases, the molar conductivity for the single cutoff case is slightly higher. For monovalent ions or roughly symmetric electrolytes, no difference is seen in Fig.~\ref{Fig11} even at very high concentrations. We note that the regime where a difference is seen is beyond the concentration regime analyzed in Fig.~\ref{Fig7}. Moreover, in this regime, our theory deviates from experimental results in the case of a single or three cutoffs, where the single cutoff approximation gives slightly better agreement to experiments.


\end{document}